# 2D Materials

PAPER

# Electrically tunable correlated domain wall network in twisted bilayer graphene

Hao-Chien Wang 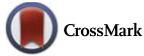 and Chen-Hsuan Hsu[*] 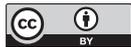

Institute of Physics, Academia Sinica, Taipei 11529, Taiwan
[*] Author to whom any correspondence should be addressed.

E-mail: chenhsuan@gate.sinica.edu.tw

## Abstract

We investigate the domain wall network in twisted bilayer graphene (TBG) under the influence of interlayer bias and screening effect from the layered structure. Starting from the continuum model, we analyze the low-energy domain wall modes within the moiré bilayer structure and obtain an analytic form representing charge density distributions of the two-dimensional structure. By computing the screened electron–electron interaction strengths both within and between the domain walls, we develop a bosonized model that describes the correlated domain wall network. We demonstrate that these interaction strengths can be modified through an applied interlayer bias, screening length and dielectric materials, and show how the model can be employed to investigate various properties of the domain wall network and its stability. We compute correlation functions both without and with phonons. Including electron–phonon coupling in the network, we establish phase diagrams from these correlation functions. These diagrams illustrate electrical tunability of the network between various phases, such as density wave states and superconductivity. Our findings reveal the domain wall network as a promising platform for the experimental manipulation of electron–electron interactions in low dimensions and the study of strongly correlated matter. We point out that our investigation not only enhances the understanding of domain wall modes in TBG but also has broader implications for the development of moiré devices.

## 1. Introduction

Low-dimensional systems offer intriguing platforms for exploring correlated electron systems. Interacting electrons in one dimension, for instance, defy the traditional quasiparticle concept, leading to the formation of Tomonaga–Luttinger liquids (TLLs) [1–4]. This framework has been extended to two-dimensional systems, giving rise to ideas of coupled-wire models. These models have been used to describe various strongly correlated systems, ranging from earlier studies on *sliding Tomonaga–Luttinger liquid* (sTLL) or *crossed sliding TLL* (csTLL) for the normal state of high-temperature superconductors [5–8] to various topologically nontrivial phases [9–21].

Remarkably, there has been a notable resurgence of interest in these coupled-wire models, largely driven by recent advancements in twisted bilayer systems [22–25], in which the twist angle between layers can be used to alter the single-particle band structures [26–34]. In particular, the domain walls in moiré twisted bilayer graphene (TBG) separating different stacking configurations have been found to host low-energy gapless modes [30, 32, 35–48], effectively forming a network of coupled quantum wires. The phenomenon is reminiscent of domain walls in Bernal-stacked bilayer graphene with opposite stacking arrangement or transverse displacement fields [49–55] and also extends beyond TBG, as similar one-dimensional channels have been identified or postulated in various nanoscale systems [56–60], such as chiral twisted trilayer graphene [61, 62], twisted WTe$_2$ [63, 64], and strain-engineered devices [65]. The discovery of one-dimensional channels across these systems has motivated theoretical exploration into effective network models [21, 66–74], and underscore the broader applicability and significance of the coupled-wire models in tunable nanoscale systems.

By identifying relevant degrees of freedom at low energy, the models provide a bosonic description for the strongly correlated twisted bilayer systems [21, 66–68, 70–72, 74], offering an alternative perspective to other theoretical works on these systems [31, 34, 75–97]. Notably, this bosonic description integrates with the renormalization-group (RG) analysis, serving as an effective tool for exploring correlated phenomena, including the correlated insulating phase and superconductivity (SC) observed in earlier studies [98, 99] and various unconventional states of matter or features in subsequent works [100–126]. While these observations are typically associated with the quasiflat bands in devices near a certain twist angle (referred to as the magic angle), here we explore a regime where the system exhibits correlation effects even when away from the magic angle.

Crucially, the moiré systems provide an opportunity to systematically determine the interaction strengths in devices consisting of one-dimensional channels. This is a notable contrast to earlier studies on sTLL and csTLL in non-moiré systems [5–8], which often relied on ad hoc assumptions regarding the specific forms of the interaction strength. Our investigation is also inspired by recent studies [49, 50, 73, 113] that have shown the tunability of interaction strength in graphene-based devices. In [113], a control of electron–electron interactions was achieved by altering the distance between TBG and a metallic screening layer, observing a suppression of correlated insulating phases in devices with a smaller screening distance. Additionally, domain wall modes in Bernal-stacked bilayer graphene can be manipulated through voltage gating [49], which subsequently alters the properties of TLL [50]. Similar domain wall modes appearing in large-angle TBG can also be electrically controlled [73]. These insights further substantiate the systematic controllability of the interaction strength in the domain wall network.

In this article, we investigate the domain wall network in TBG under the influence of interlayer bias. We compute the energy dispersion and density profile of the low-energy domain wall modes within the moiré bilayer structure. By globally fitting the charge density in the two-dimensional structure, we derive an analytical expression representing charge density distributions. This allows for efficient calculations of electron–electron interaction strengths within and between domain walls. Our investigation extends to how these interaction strengths can be modulated by externally controllable parameters, highlighting the potential for experimental manipulation through device design and preparation. Building on this model, we explore the spectroscopic and transport properties of the network, uncovering potential Anderson localization in moderately disordered devices. We also examine the instability of the network towards various density wave (DW) states and SC in two scenarios, without and with phonons. In purely electronic systems, repulsive electron–electron interactions favor the formation of charge density waves (CDW) and spin density waves (SDW). However, longitudinal acoustic phonons can enhance pairing instability within the domain wall network, potentially driving these systems towards SC. In the presence of very strong electron–phonon coupling, we identify the Wentzel–Bardeen (WB) singularity that eventually destabilize the network. Our findings establish the TBG network as an electrically tunable, low-dimensional platform for the systematic investigation of strongly correlated phenomena.

The article is structured as follows. In section 2, we introduce the continuum model in the presence of an applied interlayer bias. In section 3, we calculate the spatial charge density distribution from the continuum model and derive an analytic form via global fitting, examining the influence of external parameters on this distribution. Section 4 focuses on computing the effective interaction strength for both intrawire and interwire terms, including their dependence on interlayer bias, screening length, and dielectric materials. In section 5, drawing insights from the single-particle model, we construct an interacting model for the correlated domain wall network. Here, we investigate its spectroscopic and transport properties, along with quantifying the localization length and temperature for the Anderson localization of the network. In section 6, we examine the instability of the network towards DW and SC, either in purely electronic systems in section 6.1 or in the presence of phonons in section 6.2. Our findings and their broader implications are discussed in section 7. Appendix A provides an analysis on the influence of the chemical potential variation on the domain wall network. Additional details on the effective action and correlation functions are provided in appendix B for purely electronic systems and in appendix C in the presence of phonons, respectively.

## 2. Continuum model of TBG in the presence of interlayer bias

The continuum model of TBG, originally developed in [26], has been widely employed to study the band structure of the system at the single-particle level [29, 31, 32, 34, 62]. As a starting point, we incorporate the interlayer bias in this model [30],

$$H_{sp} = H_0 + H_{\text{hyb}}, \quad (1a)$$

with $H_0$ describing the top and bottom graphene layers with a twist angle $\theta$ and $H_{\text{hyb}}$ the hybridization

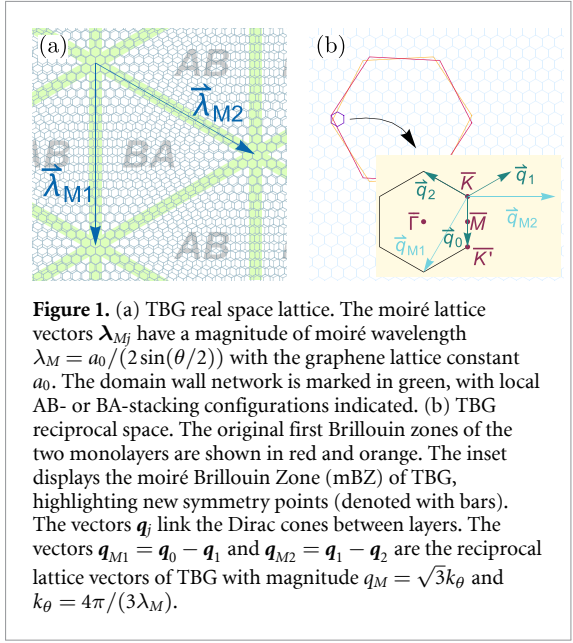

**Figure 1.** (a) TBG real space lattice. The moiré lattice vectors $\boldsymbol{\lambda}_{Mj}$ have a magnitude of moiré wavelength $\lambda_M = a_0/(2\sin(\theta/2))$ with the graphene lattice constant $a_0$. The domain wall network is marked in green, with local AB- or BA-stacking configurations indicated. (b) TBG reciprocal space. The original first Brillouin zones of the two monolayers are shown in red and orange. The inset displays the moiré Brillouin Zone (mBZ) of TBG, highlighting new symmetry points (denoted with bars). The vectors $\boldsymbol{q}_j$ link the Dirac cones between layers. The vectors $\boldsymbol{q}_{M1} = \boldsymbol{q}_0 - \boldsymbol{q}_1$ and $\boldsymbol{q}_{M2} = \boldsymbol{q}_1 - \boldsymbol{q}_2$ are the reciprocal lattice vectors of TBG with magnitude $q_M = \sqrt{3}k_\theta$ and $k_\theta = 4\pi/(3\lambda_M)$.

between the two layers. For given valley and spin, this single-particle Hamiltonian can be expressed as

$$H_{sp}\Big|_{\gamma\sigma} = \begin{pmatrix} h_D^\gamma\left(\frac{\theta}{2}\right) + V_d & T_{\text{hyb}}^\gamma \\ \left(T_{\text{hyb}}^\gamma\right)^\dagger & h_D^\gamma\left(-\frac{\theta}{2}\right) - V_d \end{pmatrix}, \quad (1b)$$

in the basis of $(c_{A\gamma\sigma}^1, c_{B\gamma\sigma}^1, c_{A\gamma\sigma}^2, c_{B\gamma\sigma}^2)^\intercal$ with the fermion field $c_{\alpha\gamma\sigma}^\eta$, the transpose operator $\intercal$, and the indices corresponding to the layer $\eta \in \{1,2\}$, sublattice $\alpha \in \{A,B\}$, valley $\gamma \in \{K \equiv +1, K' \equiv -1\}$, and spin $\sigma \in \{\uparrow,\downarrow\}$. In the above, the diagonal blocks incorporate the interlayer bias $2V_d$ and contain the following term representing the Dirac cone at the $\gamma$ valley,

$$h_D^\gamma(\theta) = v_F(\boldsymbol{p} - \hbar\boldsymbol{K}^\gamma) \cdot e^{\frac{i\gamma\theta}{2}\tau_z}\boldsymbol{\tau}^\gamma e^{-\frac{i\gamma\theta}{2}\tau_z}, \quad (2)$$

where we have the Fermi velocity $v_F$ for monolayer graphene, the momentum operator $\boldsymbol{p} = (p_x, p_y)$, and the location $\boldsymbol{K}^\gamma$ of the Dirac point in momentum space. We denote $\boldsymbol{\tau}^\gamma = (\gamma\tau^x, \tau^y)$ with the component $\mu \in \{x,y,z\}$ of the Pauli matrix $\tau^\mu$ acting on the sublattice index. The off-diagonal blocks in equation (1b) describe the spatially dependent interlayer hybridization and act as a periodic moiré potential with moiré wavelength $\lambda_M$ defined in the caption of figure 1(a). Setting the origin at one of the AA-stacking region centers, we have $T_{\text{hyb}}^\gamma = \sum_{j=0}^{2} e^{i\gamma\boldsymbol{q}_j\cdot\boldsymbol{r}}T_j^\gamma$, with the two-by-two matrix

$$T_j^\gamma = w_{AA}\tau^0 + w_{AB}\left[\tau^x\cos\left(\frac{2j\pi}{3}\right) + \gamma\tau^y\sin\left(\frac{2j\pi}{3}\right)\right]. \quad (3)$$

Here, we have $\boldsymbol{q}_j = \mathcal{R}_{2\pi j/3}(0, -k_\theta)$, with the rotation operator $\mathcal{R}_\phi$ and $k_\theta$ defined in figure 1 caption. We account for the effect of lattice relaxation by allowing for different amplitudes for hoppings between identical ($w_{AA}$) and distinct ($w_{AB}$) sublattices between the layers [29, 31, 32, 62, 127]. Following [32], we adopt an value for the ratio $w_{AA}/w_{AB}$ between 0.2 and 0.5, assuming moderate relaxation effect. As the formulations in [26, 33, 34], we define the dimensionless parameters $\alpha_{AA} = w_{AA}/(\hbar v_F k_\theta)$ and $\alpha_{AB} = w_{AB}/(\hbar v_F k_\theta)$. We will term the latter as the effective hybridization parameter, which plays a key role in the band structure.

As in [26, 32, 33, 128], we proceed by expressing the single-particle Hamiltonian in equation (1) in a plane wave basis, implementing a momentum cutoff at the scale determined by the larger of $\alpha_{AB}q_M$ and $V_d q_M/(\hbar v_F k_\theta)$ with $q_M$ specified in the caption of figure 1. The resulting expression allows us to perform numerical diagonalization. In addition to the band structure, the model introduced here allows us to compute the spatial distribution of the charge density, which we investigate next.

## 3. Charge density distribution under interlayer bias

Motivated by domain wall network discussed in [30], we explore the gapless modes in the lowest two bands near the charge neutrality point around the $K$ and $K'$ points in the original first Brillouin zone of monolayer graphene. More precisely, for a given valley (for instance, $K$), we focus on the low-energy modes around $\overline{K}$ and $\overline{K}'$ points in the mBZ, as indicated in figure 1(b). Aiming at exploring domain wall modes, we set $V_d$ to nonzero values and focus on the parameter range where $V_d/\hbar v_F k_\theta \gtrsim \alpha_{AB} > 1$ for the rest of the article.

By numerically diagonalizing the single-particle Hamiltonian, we compute the wave functions and subsequently obtain the spatial distribution of the charge density. An example is illustrated in figure 2, where we display the computed charge density alongside the corresponding state in the mBZ. It can be observed that the charge density is predominantly concentrated along the domain walls in the direction of propagation. Furthermore, within these domain walls, the density exhibits higher values around the AA-stacking regions. This latter feature aligns with findings in previous studies [31, 32], but not well described by the simple approximations in [30]. We note that a smaller twist angle leads to a shift of density from the AA-stacking regions to the domain wall segments, resulting in a more uniform distribution along the entire domain wall [43]. However, this smaller twist angle also increases $\alpha_{AB}$, thereby requiring a higher momentum cutoff in our numerical diagonalization. For illustrative purposes, we mainly use a larger twist angle $\theta = 0.5°$, unless specified otherwise.

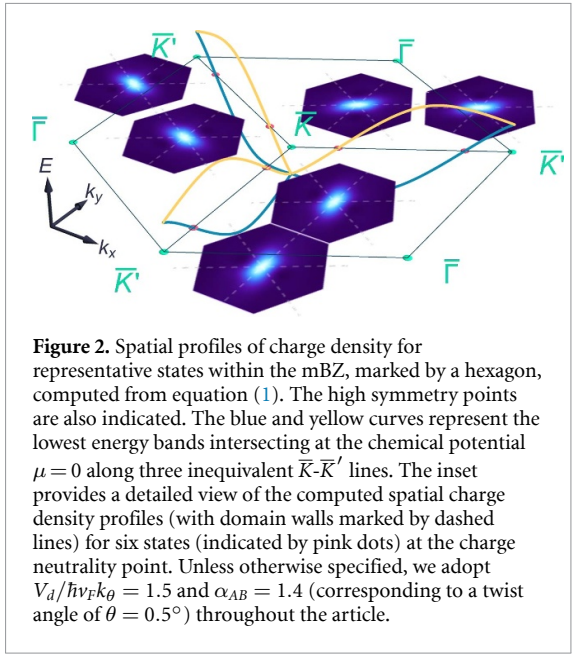

**Figure 2.** Spatial profiles of charge density for representative states within the mBZ, marked by a hexagon, computed from equation (1). The high symmetry points are also indicated. The blue and yellow curves represent the lowest energy bands intersecting at the chemical potential $\mu = 0$ along three inequivalent $\bar{K}$-$\bar{K}'$ lines. The inset provides a detailed view of the computed spatial charge density profiles (with domain walls marked by dashed lines) for six states (indicated by pink dots) at the charge neutrality point. Unless otherwise specified, we adopt $V_d/\hbar v_F k_\theta = 1.5$ and $\alpha_{AB} = 1.4$ (corresponding to a twist angle of $\theta = 0.5°$) throughout the article.

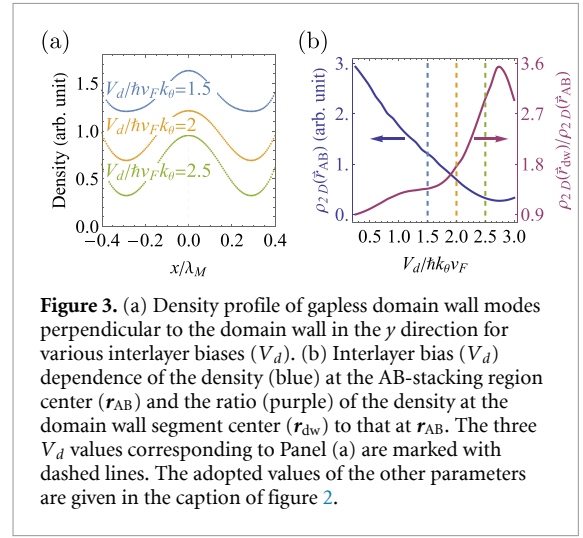

**Figure 3.** (a) Density profile of gapless domain wall modes perpendicular to the domain wall in the $y$ direction for various interlayer biases ($V_d$). (b) Interlayer bias ($V_d$) dependence of the density (blue) at the AB-stacking region center ($\boldsymbol{r}_{AB}$) and the ratio (purple) of the density at the domain wall segment center ($\boldsymbol{r}_{dw}$) to that at $\boldsymbol{r}_{AB}$. The three $V_d$ values corresponding to Panel (a) are marked with dashed lines. The adopted values of the other parameters are given in the caption of figure 2.

As depicted in the main panel of figure 2, in the momentum space, for each valley and spin, there are two energy band branches hosting these low-energy modes. Reversing the valley inverts the velocity of the modes, indicating electrons propagating in the opposite direction. This feature is consistent with findings in TBG [30] and also noted in nontwisted samples [52–54]. The charge density profiles presented in the insets of figure 2 correspond to $\mu = 0$. We have verified that the results remain largely unchanged when varying the chemical potential, provided it stays within the narrow bands depicted in the main panel. A more detailed analysis on the effects of chemical potential on the density distributions of the domain wall modes is provided in appendix A. It is important to note that our findings apply to systems with a rather small twist angle, which contrasts with systems near the magic angle, where the chemical potential can induce nonnegligible effects [85, 129–136].

To resolve the domain wall modes more clearly, we focus on one of the domain walls along the $y$ direction. In figure 3(a), we show the spatial dependence of the charge density perpendicular to the domain wall for several $V_d$ values. In addition to confirming the confinement at the domain wall, we observe that the density profile can be electrically modified. Namely, as the interlayer bias increases, the charge density becomes more confined within the domain wall, as a result of the local spectral gap induced by the bias at Bernal-stacking domains [30, 137, 138]. This feature can be examined more quantitatively. In figure 3(b), we plot the density at the center of the AB-stacking region (labeled by $\boldsymbol{r}_{AB}$) as a function of the interlayer bias, observing a decrease in density at $\boldsymbol{r}_{AB}$ due to the increasing local gap with the interlayer bias. To illustrate the density shift with respect to the interlayer bias, we also show the ratio of the density at the center of the domain wall segment (labeled by $\boldsymbol{r}_{dw}$) to that at $\boldsymbol{r}_{AB}$. As the local gap within the domain regions is enhanced by increasing interlayer bias, the reduction in the density at $\boldsymbol{r}_{AB}$ leads to increase in the density at $\boldsymbol{r}_{dw}$. Since the interactions within the domain wall network are governed by the screened Coulomb potential, which involves essentially short-range density–density interactions, this feature suggests the tunability of these interactions through external electrical control.

In addition to locating the low-energy modes in real space and determining their spatial profile, we are also able to extract the bandwidth $\Delta_a$ and the Fermi velocity $v_{dw}$ of the domain wall modes through the dispersions of domain wall modes in the spectrum. We summarize the results in figure 4, where we show how the two quantities influenced by the interlayer bias and effective hybridization parameter (quantified by $\alpha_{AB}$). Again, the dependence on the interlayer bias demonstrates the electrical tunability of the domain wall network. As $V_d$ increases, the bands around the charge neutrality point flatten, and the bandwidth $\Delta_a$ decreases, a feature also observed in [45]. Crucial for our subsequent analysis, this leads to a decreased $v_{dw}$ value. On the other hand, the dependence on the effective hybridization parameter indicates variations in these quantities across different samples, though with a weaker dependence in the studied range. Specifically, since $\alpha_{AB}$ is influenced by the hybridization strength $w_{AB}$ and the moiré wavelength $\lambda_M$, the properties of the network can be altered through device preparation, either by adjusting $w_{AB}$ through applied pressure (experimentally achieved in [106]) or by varying $\lambda_M$ through the twist angle.

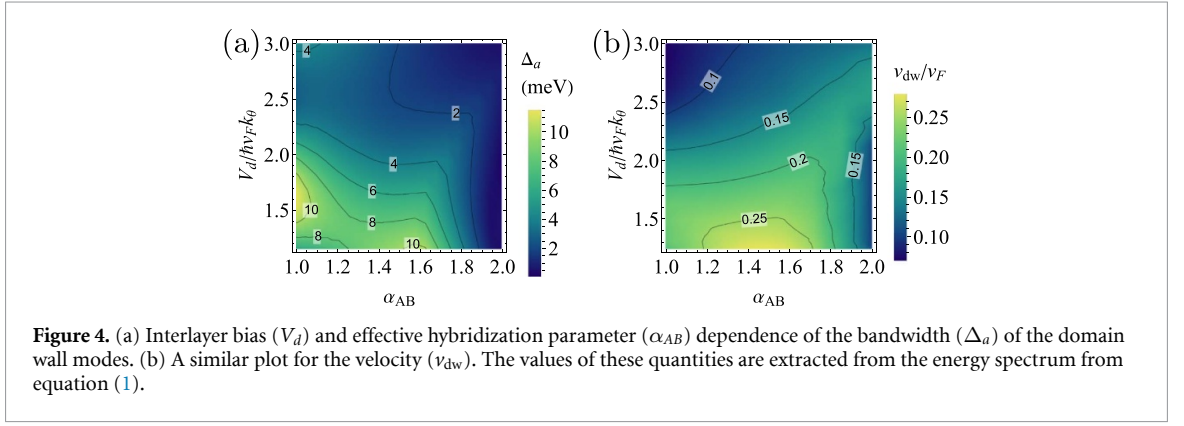

**Figure 4.** (a) Interlayer bias ($V_d$) and effective hybridization parameter ($\alpha_{AB}$) dependence of the bandwidth ($\Delta_a$) of the domain wall modes. (b) A similar plot for the velocity ($v_{dw}$). The values of these quantities are extracted from the energy spectrum from equation (1).

Having established the presence of domain wall modes, we introduce a fitting function in order to express the density using an analytic form, which will be useful when we estimate the interaction strength below. Specifically, we take the following functional form,

$$\rho_{2D}(\boldsymbol{r}) = C_0 \sum_{j=0}^{2} \sum_{m} \sum_{\delta} \rho_{\delta,m}^{dw}(\mathcal{R}_{2j\pi/3}\boldsymbol{r}), \quad (4a)$$

$$\rho_{\delta,m}^{dw}(\boldsymbol{r}) = \rho_{\perp\delta,m}(x)\rho_{\|\delta,m}(y)\rho_z(z), \quad (4b)$$

$$\rho_{\perp\delta,m}(x) = e^{-8\kappa_\perp \sin^2\left[\frac{\sqrt{3}}{4}k_\theta\left(x-\frac{\sqrt{3}}{2}m\lambda_M\right)\right]}$$
$$\times \Theta\left(x+\sqrt{3}(1+2m)\lambda_M/4\right)$$
$$\times \Theta\left(\sqrt{3}(1+2m)\lambda_M/4 - x\right),$$

$$\rho_{\|\delta,m}(y) = \left\{e^{-8\kappa_\| \sin^2\left[\frac{3}{4}k_\theta\left(y-\frac{m}{2}\lambda_M\right)\right]} + c_{0\|}\right\}$$
$$\times \Theta\left(y+L_\|/2\right)\Theta\left(L_\|/2 - y\right),$$

$$\rho_z(z) = \Theta(z)\Theta(L_z - z),$$

with the normalization constant $C_0$, domain wall length $L_\|$, the extent $L_z$ of the wave function perpendicular to the layered structure, the Heaviside function $\Theta(x)$ and the index $m$ labeling the domain wall for a given domain wall direction indicated by $j$. Here, we define the index[1] $\delta \in \{1,2\}$ to denote the branches of the gapless modes in the spectrum, which are represented by the blue and yellow curves in figure 2.

In the above, we adopt a functional form that allows for the separation of the two orthogonal directions locally defined through the domain wall orientation. The parameter $\kappa_\perp$ is introduced to characterize the confinement perpendicular to the domain wall. The form of $\rho_\perp$ is inspired by the Jackiw–Rebbi solution as described in [30]. Partially motivated by the overall $C_{3z}$ symmetry of TBG [33], we adopt an empirical formula $\rho_\|$ of a form similar to the transverse component to describe the density profile along the domain wall, with a higher density near the AA-stacking regions characterized by the parameter $\kappa_\|$. In the above expression, $\kappa_\perp$, $\kappa_\|$ and $c_{0\|}$ serve as the fitting parameters, which can be determined by globally fitting[2] the two-dimensional density obtained from the continuum model to $\rho_{2D}(\boldsymbol{r})$.

An example of such fitting is illustrated in figure 5, where we show the computed density profile (blue dots) and the fitting results (red curves) along two directions, one transverse to the wall (in the $x$ direction), and the other along the domain wall (in the $y$ direction). We see that the expression $\rho_\|$ along the domain wall fits well, including the peaks around the AA-stacking regions. Perpendicular to the domain wall, the fit is reasonably good, although some discrepancies are observed. It is important to note that these discrepancies are of minor significance in the two-dimensional global fitting due to the smaller overall magnitude compared to the AA-stacking centers.

Through our fitting procedure, we establish how the system parameters influence the fitted values, as illustrated in figure 6. The local spectral gap in the Bernal-stacking regions, which increases with the interlayer bias $V_d$ [137, 138], leads to a corresponding increase in both $\kappa_\perp$ and $\kappa_\|$ (approximately linearly). This observation further confirms the enhanced confinement within the domain wall and AA-stacking regions. Consequently, these findings support the idea that it is possible to manipulate charge properties and interaction strength within the domain walls using externally controllable parameters.

---

[1] However, our numerical analysis indicates no practical differences in the spatial profile of $\rho_{\delta,m}^{dw}$ between the branches $\delta = 1$ and $\delta = 2$. Consequently, we refrain from introducing redundant labels for the parameters $\kappa_\perp$, $\kappa_\|$, and $c_{0\|}$. Therefore, the primary distinction between the two branches lies in their wave functions in momentum space, which will be distinguished using different Fermi momenta in the subsequent sections.

[2] Here, we introduce a method to expedite the calculation of the integral for density–density interaction strength below. It is possible to use a different functional form for fitting. As an alternative, the interaction strength can also be computed numerically through brute-force calculation. In any case, these choices do not alter our conclusion.

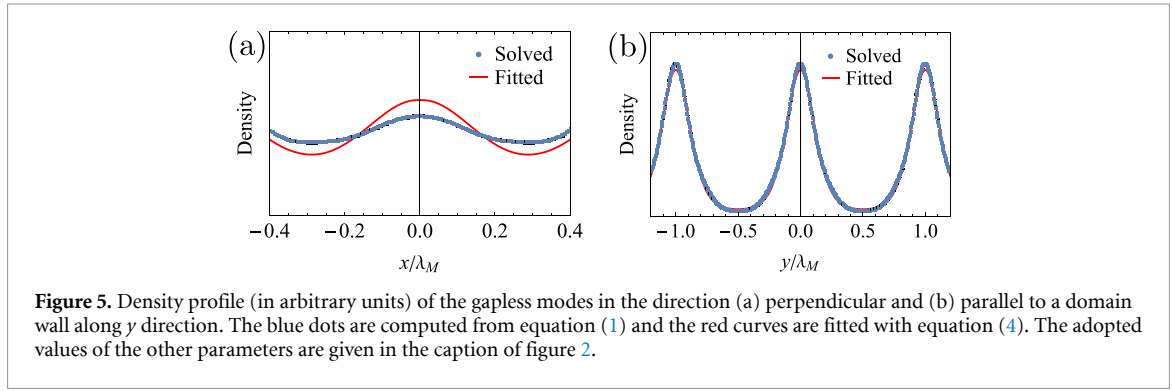

**Figure 5.** Density profile (in arbitrary units) of the gapless modes in the direction (a) perpendicular and (b) parallel to a domain wall along *y* direction. The blue dots are computed from equation (1) and the red curves are fitted with equation (4). The adopted values of the other parameters are given in the caption of figure 2.

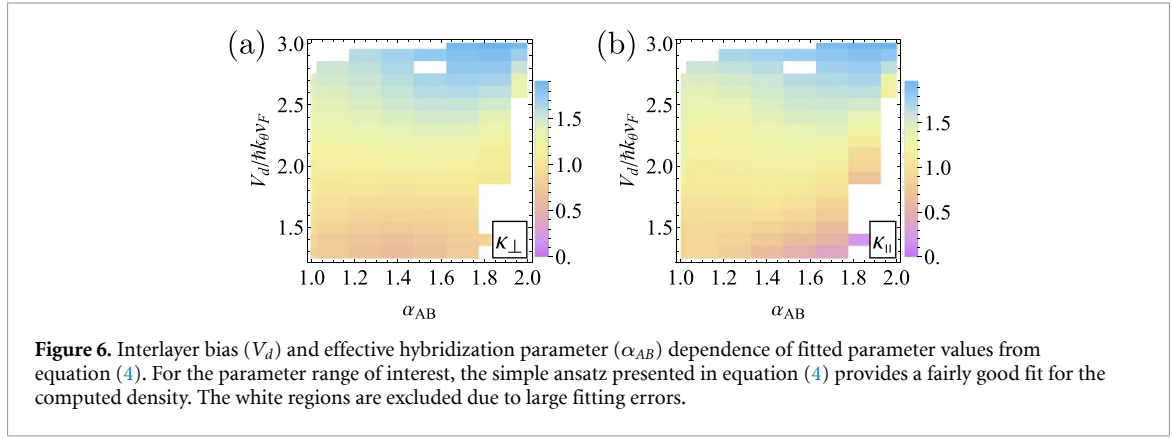

**Figure 6.** Interlayer bias ($V_d$) and effective hybridization parameter ($\alpha_{AB}$) dependence of fitted parameter values from equation (4). For the parameter range of interest, the simple ansatz presented in equation (4) provides a fairly good fit for the computed density. The white regions are excluded due to large fitting errors.

## 4. Interaction strength in the domain wall network

With the analytic form of the spatial density profile and its parameters deduced from the fitting procedure, we now compute the effective interaction strength for both intrawire (within a domain wall) and interwire (between parallel domain walls) contributions. The empirical formula in equation (4) allows for a straightforward separation of variables parallel and perpendicular to the domain walls. To proceed, we analyze a device configuration where the TBG is positioned on top of a dielectric layer with thickness $d$ (setting $z = 0$ at the interface). This layered structure is then situated above a metallic back gate, located in the region $z \leqslant -d$, as depicted in figure 7. The presence of the metallic gate results in a screened Coulomb interaction, making it effectively short-ranged. Consequently, the thickness of the dielectric layer serves as the screening length, quantifying the degree of the screening effect.

In general, screening effects in such a three-layer structure can be formally incorporated through an infinite sum of image charges [140, 141]. Here, since in our setup the electron wave function in the TBG has a narrow distribution in the direction perpendicular to the layered structure (more quantitatively, $L_z \ll d$), we simplify our problem by considering an image charge distribution (caused by the charges in TBG) in $-L_z \leqslant z \leqslant 0$ with a screening factor of $-a_{\text{scn}}$ [142] and another distribution (caused by the TBG and dielectric layer) in the metallic layer at $-2d - L_z \leqslant z \leqslant -2d$, with the charge multiplied by $-(1 - a_{\text{scn}})$. Here, we introduce the parameter, $a_{\text{scn}} = (\epsilon_r - 1)/(\epsilon_r + 1)$, with the relative dielectric constant $\epsilon_r$. With this approximation, for the distribution $\rho_{\delta,m}^{\text{dw}}$ of the branch $\delta$ in a given domain wall labeled by $m$, we have the following expressions for the image charge density,

$$\rho_{\delta,m}^{\text{diel}}(x,y,z) = -a_{\text{scn}} \rho_{\delta,m}^{\text{dw}}(x,y,-z), \quad (5a)$$

$$\rho_{\delta,m}^{\text{met}}(x,y,z) = -(1 - a_{\text{scn}}) \rho_{\delta,m}^{\text{dw}}(x,y,-z-2d). \quad (5b)$$

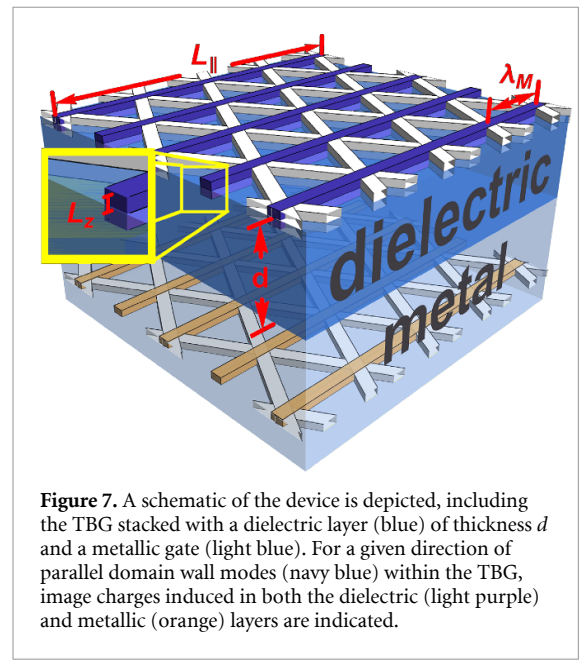

**Figure 7.** A schematic of the device is depicted, including the TBG stacked with a dielectric layer (blue) of thickness $d$ and a metallic gate (light blue). For a given direction of parallel domain wall modes (navy blue) within the TBG, image charges induced in both the dielectric (light purple) and metallic (orange) layers are indicated.

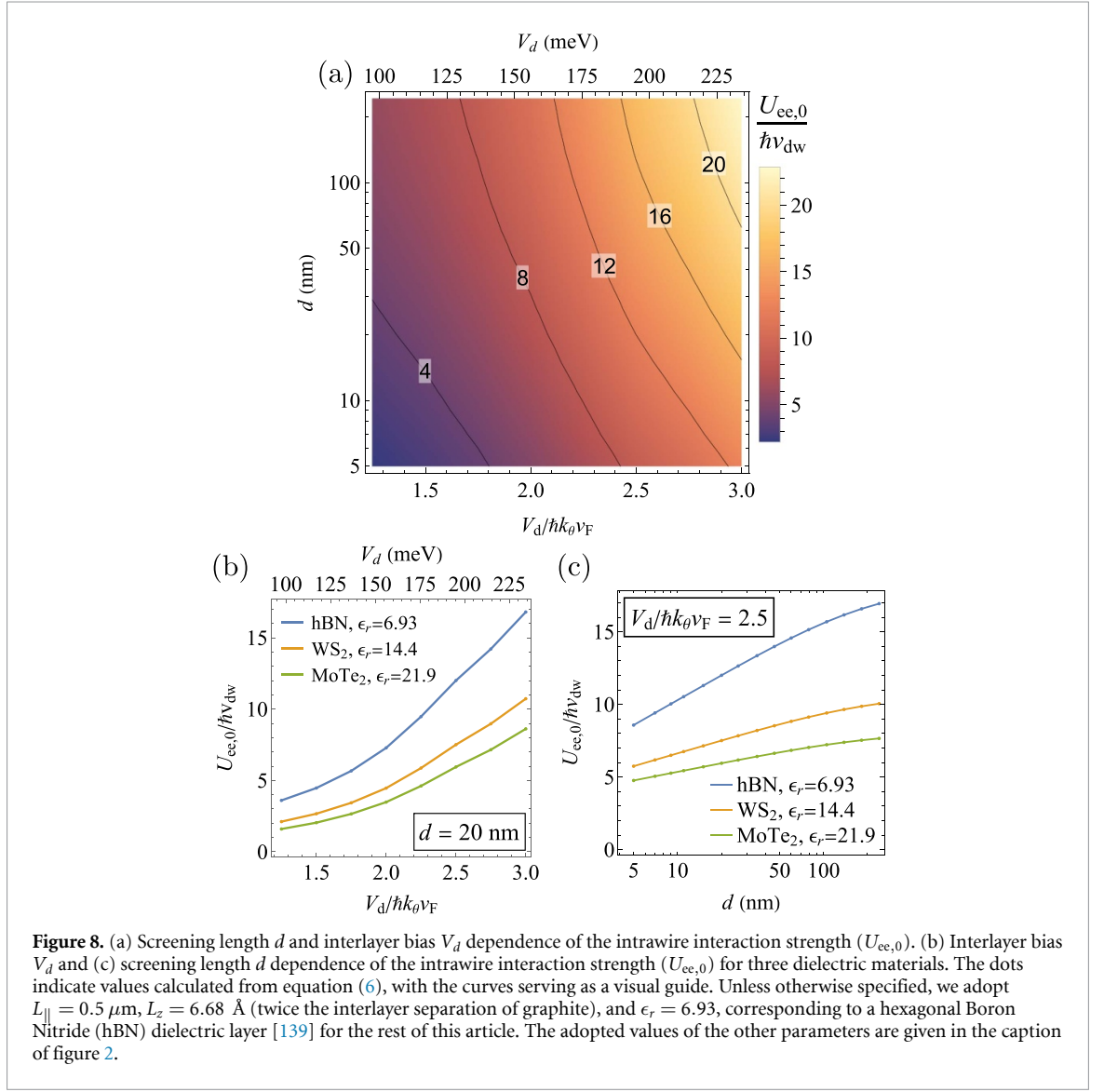

**Figure 8.** (a) Screening length $d$ and interlayer bias $V_d$ dependence of the intrawire interaction strength ($U_{ee,0}$). (b) Interlayer bias $V_d$ and (c) screening length $d$ dependence of the intrawire interaction strength ($U_{ee,0}$) for three dielectric materials. The dots indicate values calculated from equation (6), with the curves serving as a visual guide. Unless otherwise specified, we adopt $L_{\parallel} = 0.5\,\mu$m, $L_z = 6.68$ Å (twice the interlayer separation of graphite), and $\epsilon_r = 6.93$, corresponding to a hexagonal Boron Nitride (hBN) dielectric layer [139] for the rest of this article. The adopted values of the other parameters are given in the caption of figure 2.

The intrabranch and interbranch electrostatic energy between the $m$th domain wall and $(m+n)$th domain wall can be expressed as

$$U_{ee,n} = \frac{e^2 L_{\parallel}}{4\pi\epsilon_0} \sum_{\mathfrak{M}} \int d\mathbf{r} \int d\mathbf{r}' \frac{\rho_{\delta,m}^{\text{dw}}(\mathbf{r}) \rho_{\overline{\delta},m+n}^{\mathfrak{M}}(\mathbf{r}')}{|\mathbf{r}-\mathbf{r}'|}, \quad (6a)$$

$$V_{ee,n} = \frac{e^2 L_{\parallel}}{4\pi\epsilon_0} \sum_{\mathfrak{M}} \int d\mathbf{r} \int d\mathbf{r}' \frac{\rho_{\delta,m}^{\text{dw}}(\mathbf{r}) \rho_{\overline{\delta},m+n}^{\mathfrak{M}}(\mathbf{r}')}{|\mathbf{r}-\mathbf{r}'|}, \quad (6b)$$

where $\mathfrak{M} \in \{\text{dw, diel, met}\}$ denotes different contributions and $\overline{\delta}$ denotes the branch opposite to $\delta$. As mentioned earlier, as we do not find numerical difference between $\rho_{1,m}^{\text{dw}}$ and $\rho_{2,m}^{\text{dw}}$, the above integrals are practically the same for any $\delta$. To proceed, we reformulate the denominator in the integrand of equation (6) using a Gaussian integral [143–145]. Using this approach, the integral over $z$ becomes straightforward owing to the uniform distribution in this direction. In contrast, the integrals over the in-plane spatial coordinates will be evaluated numerically.

The main results of this numerical procedure are summarized in figures 8 and 9, for the intrawire ($U_{ee,n=0}$) and interwire ($U_{ee,n\neq 0}$) terms, respectively. For the intrawire interaction strength in figure 8, we observe an increase with a larger interlayer bias due to the stronger confinement of the electron wave functions in the domain walls, consistent with the computed density profile in section 3. In addition, there is a decrease in $U_{ee,0}$ with a larger dielectric constant and a shorter distance from the metallic layer owing to the stronger screening effect. Notably, the interaction strength can be enhanced by *more than seven fold* in the investigated parameter regime, demonstrating great electrical tunability.

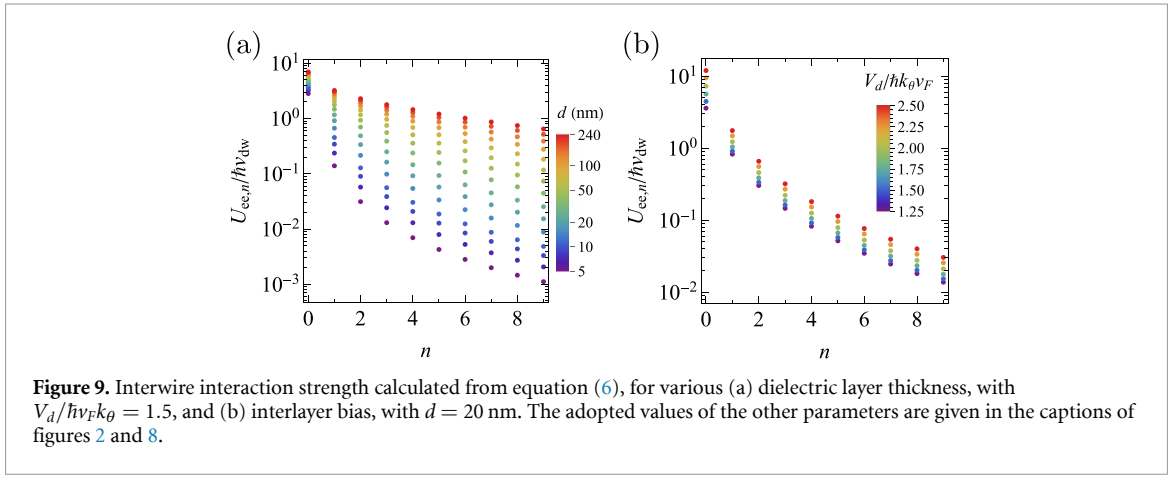

**Figure 9.** Interwire interaction strength calculated from equation (6), for various (a) dielectric layer thickness, with $V_d/\hbar v_F k_\theta = 1.5$, and (b) interlayer bias, with $d = 20$ nm. The adopted values of the other parameters are given in the captions of figures 2 and 8.

As to the interaction strength between electrons in parallel domain walls, we have calculated the interwire terms up to ninth neighboring domain walls, as shown in figure 9. The dependence of $U_{ee,n}$ for a given $n$ on interlayer bias and screening length follows the same trend as the intrawire term. For different $n$, while the interwire interaction is more severely screened by a shorter $d$ for domain walls that are farther apart (see figure 9(a)), the $V_d$ dependence behaves similarly (see figure 9(b)). The quantitative differences observed between the two panels are expected. Specifically, the interaction between the $n$th nearest neighbor domain walls is significantly reduced when the screening length is comparable to their separation distance (that is, $d \sim \sqrt{3}n\lambda_M/2$). This reduction in interaction strength is less significant when $d$ is much larger. On the other hand, the dependence of the interaction strength on the bias voltage originates from the confinement of electron wave functions induced by $V_d$. Consequently, this control parameter influences the strength uniformly across different values of $n$, resulting in a minor dependence on the latter. Since the strength of the interwire interaction decays with the separation between domain walls, as expected, it allows for the exclusion of interwire interaction terms at large $n$ in the construction of the interacting network model. The rate of decay changes with the distance $d$ from the metallic gate, with a smaller $d$ value leading to a faster decay.

Before concluding this section, we note previous experimental studies that have demonstrated tuning the interaction strength in Bernal-stacked bilayer graphene through $V_d$ [50], as well as in unbiased TBG by varying $d$ [113]. With the demonstrated tunability here, we expect future efforts towards systematic investigations in biased TBG, either by utilizing the aforementioned control methods or by exploring the use of various dielectric materials. Beginning with the continuum model, we have analyzed the charge density and the effective interaction strengths within the domain wall network. This analysis provides the foundation for constructing our model describing interacting electrons in the network, which we introduce in the following section.

## 5. Correlated domain wall network

With the interaction strengths within the network determined, we are set to establish an interacting model for the domain wall modes. This model distinctively considers two branches of gapless modes for each direction of motion and therefore differs from the previous bosonization models for moiré networks [21, 67, 68, 72], which neglected these additional degrees of freedom. With our description, each domain wall is reminiscent of the bosonization model for metallic carbon nanotubes [146–148], where the doubling in degrees of freedom is due to the two Dirac cones in the spectrum. Here, the situation is further complicated by the formation of the moiré network, leading our model to effectively describe interacting electrons in coupled nanotubes forming a mesoscopic network.

To be explicit, we describe electrons as fermion fields $\psi^{(j)}_{\sigma,m}$ in each domain wall in figure 10 and bosonize them as follows,

$$\psi^{(j)}_{\sigma,m}(x) = \sum_{\ell\delta} \psi^{(j)}_{\ell\delta\sigma,m}(x),$$

$$\psi^{(j)}_{\ell\delta\sigma,m}(x) = \frac{U^j_{\ell\delta\sigma,m}}{\sqrt{2\pi a}} e^{i\ell k^{(j)}_{F\delta,m} x} \exp\left\{\frac{i}{2}\Big[-\ell\left(\phi^j_{cs,m} + \delta\phi^j_{ca,m}\right)\right.$$
$$-\ell\sigma\left(\phi^j_{ss,m} + \delta\phi^j_{sa,m}\right) + \left(\theta^j_{cs,m} + \delta\theta^j_{ca,m}\right)$$
$$\left.+ \sigma\left(\theta^j_{ss,m} + \delta\theta^j_{sa,m}\right)\Big]\right\}, \quad (7)$$

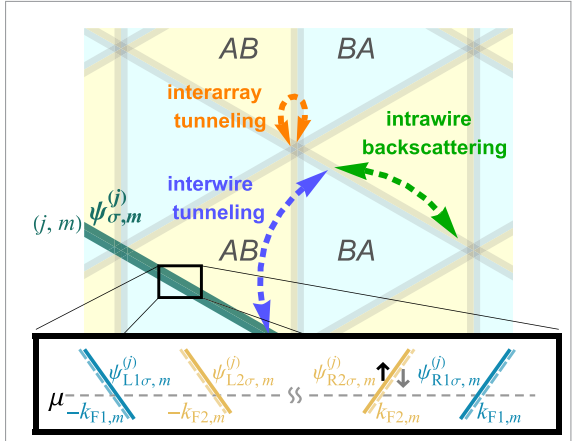

**Figure 10.** Illustration of various microscopic processes in the domain wall network discussed in section 5. The inset shows the linearized energy spectrum near the Fermi level in the $m$th domain wall of the $j$th array. Blue and yellow indicate the two branches shown in figure 2, with solid and dashed lines representing opposite spin states. The operator $\psi^{(j)}_{\ell\delta\sigma,m}$ represents a fermion field. For $\ell = R$ ($\ell = L$), it corresponds to the right- (left-) moving mode. The outer (inner) branch is denoted by $\delta = 1$ ($\delta = 2$), and $\sigma$ indicates the spin state.

with the array index $j \in \{0,1,2\}$ corresponding to the three $\boldsymbol{q}_j$ directions introduced in figure 1 caption, the domain wall index $m \in [1, N_\perp]$, the moving-direction index $\ell \in \{R \equiv +, L \equiv -\}$ along a given domain wall, the spin index $\sigma \in \{\uparrow \equiv +, \downarrow \equiv -\}$, the local coordinate $x$, the Fermi wave vector $k^{(j)}_{F\delta,m}$, and $\delta \in \{1 \equiv +, 2 \equiv -\}$ labeling the outer/inner branch of the domain wall spectrum. In the bosonic description, we introduce the Klein factor $U^j_{\ell\delta\sigma,m}$, the short-distance cutoff set by $a = \hbar v_{\rm dw}/\Delta_a$, and the boson fields satisfying

$$\left[\phi^j_{\nu P, m}(x), \theta^{j'}_{\nu' P', m'}(x')\right] = \frac{i\pi}{2} \delta_{jj'} \delta_{\nu\nu'} \delta_{PP'} \delta_{mm'}$$
$$\times {\rm sign}\,(x'-x), \quad (8)$$

with the index $\nu \in \{c, s\}$ labeling the charge/spin sector and $P \in \{s, a\}$ for the symmetric/antisymmetric combinaiton of the two branches. The above relation implies that $\phi^j_{\nu P, m}$ and $\partial_x \theta^j_{\nu P, m}$ are conjugate to each other.

In terms of the boson fields, our model for the correlated domain wall network takes the form,

$$H_{\rm ee} = \sum_{j=0}^{2} \sum_{P \in \{s,a\}} \left(H^{(j)}_{cP} + H^{(j)}_{sP}\right), \quad (9a)$$

$$H^{(j)}_{cP} = \sum_{m}\sum_{n}\int \frac{dx}{2\pi}\left[U^{(j)}_{\phi_{cP},n}\left(\partial_x\phi^j_{cP,m}\right)\left(\partial_x\phi^j_{cP,m+n}\right)\right.$$
$$\left. + U^{(j)}_{\theta_{cP},n}\left(\partial_x\theta^j_{cP,m}\right)\left(\partial_x\theta^j_{cP,m+n}\right)\right], \quad (9b)$$

$$H^{(j)}_{sP} = \sum_{m}\int \frac{\hbar dx}{2\pi}\left[\frac{u_{sP}}{K_{sP}}\left(\partial_x\phi^j_{sP,m}\right)^2\right.$$

$$\left.+ u_{sP}K_{sP}\left(\partial_x\theta^j_{sP,m}\right)^2\right], \quad (9c)$$

where $U^{(j)}_{\phi_{cP},n}$ and $U^{(j)}_{\theta_{cP},n}$ quantify the strengths of the density–density and current–current interaction terms between the $n$th neighbor domain walls, respectively. Their specific forms are given by

$$U^{(j)}_{\phi_{cs},n} = \hbar v_{\rm dw}\delta_{n0} + \frac{2}{\pi}\left[U^{(j)}_{{\rm ee},n} + V^{(j)}_{{\rm ee},n}\right], \quad (10a)$$

$$U^{(j)}_{\phi_{ca},n} = \hbar v_{\rm dw}\delta_{n0} + \frac{2}{\pi}\left[U^{(j)}_{{\rm ee},n} - V^{(j)}_{{\rm ee},n}\right], \quad (10b)$$

$$U^{(j)}_{\theta_{cs},n} = U^{(j)}_{\theta_{ca},n} = \hbar v_{\rm dw}\delta_{n0}, \quad (10c)$$

with $U^{(j)}_{{\rm ee},n}$ and $V^{(j)}_{{\rm ee},n}$ computed from equation (6). Our numerical analysis shows $U^{(j)}_{{\rm ee},n} \approx V^{(j)}_{{\rm ee},n}$ within the relevant parameter regime, suggesting that the charge antisymmetric sector (labeled by 'ca') is essentially noninteracting. However, we retain its notation for generality. Finally, for the spin sector, we have identical velocity $u_s = v_{\rm dw}/K_s$ and interaction parameter $K_s$ for all the domain walls.

While our description above preserves the $C_{3z}$ rotational symmetry, leading to an independence from the $j$ index, we retain this label for generality. This notation can be useful in scenarios where spatial inhomogeneity or anisotropy arises, such as from disorder or spontaneously broken symmetry. On the other hand, without imposing the translational symmetry in the direction perpendicular to the domain walls, there is a visible dependence on $m$, which we will demonstrate in the following sections. The application of bosonization in our model enables the computation of physical quantities nonperturbatively in the interaction strength.

To proceed, we introduce orthogonal matrices $\mathbf{M}^{(j)}_{\phi_{cP}}$ and $\mathbf{M}^{(j)}_{\theta_{cP}}$ that diagonalize the charge sector in equation (9b), leading to

$$H^{(j)}_{cP} = \sum_{m}\int \frac{dx}{2\pi}\left[\widetilde{U}^{(j)}_{\phi_{cP},m}\left(\partial_x\widetilde{\phi}^j_{cP,m}\right)^2\right.$$
$$\left.+ \widetilde{U}^{(j)}_{\theta_{cP},m}\left(\partial_x\widetilde{\theta}^j_{cP,m}\right)^2\right], \quad (11)$$

where the interaction strength and fields in the new basis read

$$\widetilde{U}^{(j)}_{\phi_{cP},m} = \sum_{pp'}\left(\mathbf{M}^{(j)}_{\phi_{cP}}\right)_{m,p} U^{(j)}_{\phi_{cP},|p-p'|} \left[\left(\mathbf{M}^{(j)}_{\phi_{cP}}\right)^{-1}\right]_{p',m},$$
$$(12a)$$

$$\widetilde{\phi}^j_{cP,m} = \sum_{p}\left(\mathbf{M}^{(j)}_{\phi_{cP}}\right)_{m,p} \phi^j_{cP,p}, \quad (12b)$$

and similarly for $\widetilde{U}^{(j)}_{\theta_{cP},m}$ and $\widetilde{\theta}^j_{cP,m}$. It is straightforward to check that in the new basis the fields $\widetilde{\phi}^j_{cP,m}$ and $\widetilde{\theta}^j_{cP,m}$ follow the same commutation relation as equation (8), provided that we have $\mathbf{M}^{(j)}_{\phi_{cP}} = \mathbf{M}^{(j)}_{\theta_{cP}}$. For the sake of notational clarity, we will continue to use the distinct notations in the following discussion. It should be noted that the stability of the model presented in equation (9) is ensured by the positive definiteness of the interaction matrix, which can be inferred from equation (9*b*). Additionally, this stability can be confirmed by verifying that all values of $\widetilde{U}^{(j)}_{\phi_{cP},m}$ in equation (12*a*) are positive. Indeed, we have numerically verified that these criteria are met across the entire range of parameters we have examined in this work. In section 6.2, however, it is shown that including phonons in the system can lead to a scenario where strong electron–phonon coupling destabilizes the model, resulting in the WB singularity.

Having established the bosonic model for our correlated domain wall network, we now explore its properties. To this end, we investigate physical quantities, including the local density of states (DOS), impurity-induced conductance correction, and tunneling current between parallel domain walls or nonparallel domain walls at intersections. In addition, we look into the Anderson localization induced by potential disorder in these domain walls.

### 5.1. Local DOS

The local DOS at the position $x$ in the $m$th domain wall of the $j$th array can be computed by generalizing the calculation for one-dimensional systems [149, 150]. Keeping the forward scattering contributions, which give the signature universal scaling behavior, we have

$$\rho^{(j)}_{\text{dos},m}(E) = \frac{1}{\pi} \text{Re} \left[ \int_0^\infty dt\, e^{iEt/\hbar} \sum_{\ell\delta\sigma} \right.$$
$$\left. \times \left\langle \psi^{(j)}_{\ell\delta\sigma,m}(x,t) \left[ \psi^{(j)}_{\ell\delta\sigma,m}(x,0) \right]^\dagger \right\rangle_{\text{ee}} \right], \tag{13}$$

where we define $\langle \cdots \rangle_{\text{ee}}$ with respect to the effective action in equation (B.1). With the bosonization formula in equation (7), we compute the local DOS, which can be expressed as a function of energy $E$ and temperature $T$,

$$\rho^{(j)}_{\text{dos},m}(E,T) \propto T^{\beta^{(j)}_m} \cosh\left( \frac{E}{2k_B T} \right)$$
$$\times \left| \Gamma\left( \frac{1+\beta^{(j)}_m}{2} + \frac{iE}{2\pi k_B T} \right) \right|^2, \tag{14}$$

with the Boltzmann constant $k_B$ and the interaction-dependent parameters defined as

$$\beta^{(j)}_m = \frac{1}{8} \left( \Delta^{(j)}_{\phi_{cs},m} + \Delta^{(j)}_{\theta_{cs},m} + \Delta^{(j)}_{\phi_{ca},m} + \Delta^{(j)}_{\theta_{ca},m} + K_{ss} \right.$$
$$\left. + K_{sa} + 1/K_{ss} + 1/K_{sa} \right) - 1, \tag{15a}$$

$$\Delta^{(j)}_{\phi_{cP},m} = \sum_p \sqrt{\frac{\widetilde{U}^{(j)}_{\theta_{cP},p}}{\widetilde{U}^{(j)}_{\phi_{cP},p}}} \left[ \left( \mathbf{M}^{(j)}_{\phi_{cP}} \right)_{p,m} \right]^2, \tag{15b}$$

$$\Delta^{(j)}_{\theta_{cP},m} = \sum_p \sqrt{\frac{\widetilde{U}^{(j)}_{\phi_{cP},p}}{\widetilde{U}^{(j)}_{\theta_{cP},p}}} \left[ \left( \mathbf{M}^{(j)}_{\theta_{cP}} \right)_{p,m} \right]^2. \tag{15c}$$

We see that, due to the coupling between the domain walls, the local DOS of a given domain wall depends on the intrawire and interwire interactions of the entire network. The local DOS in equation (14) follows a universal scaling curve and serves as spectroscopic features that can be verified through scanning tunneling spectroscopy as illustrated in figure 11(c). It can be checked that, at low $T$, the formula reduces to a power law $\rho^{(j)}_{\text{dos},m}(E) \propto |E|^{\beta^{(j)}_m}$. It is noteworthy that in the noninteracting limit, both $\Delta^{(j)}_{\phi_{cP},m}$ and $\Delta^{(j)}_{\theta_{cP},m}$ approach unity, while $\beta^{(j)}_m$ tends towards zero. These deviations thus serve as an experimentally accessible metric for quantifying the correlation in the network.

To understand how system parameters influence the above exponents, we calculate their dependence on the screening length and interlayer bias for a domain wall at the center of a mesoscopic device, and present the results in figure 11. Our numerical analysis reveals significant deviations from unity for both $\Delta^{(j)}_{\phi_{cP},m}$ and $\Delta^{(j)}_{\theta_{cP},m}$ in most regions of the plot, highlighting the presence of substantial correlation effects that can be observable through spectroscopic probes. As anticipated, the deviations become more pronounced with increased screening length and interlayer bias. This is particularly evident in the dependence of interlayer bias, which we attribute to the substantial reduction in $v_{\text{dw}}$ with increasing interlayer bias (see figure 4).

Besides the domain wall located near the center of a device, we extend our analysis to various domain walls denoted by different $m$. We focus on the interlayer bias dependence, as depicted in figure 12. Notably, since our approach does not assume translational invariance in the direction perpendicular to the domain walls, it leads to weak but visible quantitative differences between domain walls located at the edges and those in the interior of the two-dimensional system. We observe that both $\Delta^{(j)}_{\phi_{cP},m}$ and $\Delta^{(j)}_{\theta_{cP},m}$ show increased deviations from unity near the sample edges. Additionally, we explore the impact of the chemical potential $\mu$ on $\Delta^{(j)}_{\phi_{cP},m}$ and $\Delta^{(j)}_{\theta_{cP},m}$, finding a relatively weak dependence (see figure A1 for $m = 9$). The results indicate the robustness of the exponents (including those discussed below) with respect to the doping level, provided that the chemical potential remains within the low-energy bands under investigation. We conclude this subsection by noting that a systematic investigation of the local DOS at domain wall segments (away from the AA-stacking regions) using scanning tunneling spectroscopy could be highly useful in quantifying correlations within the domain wall network.

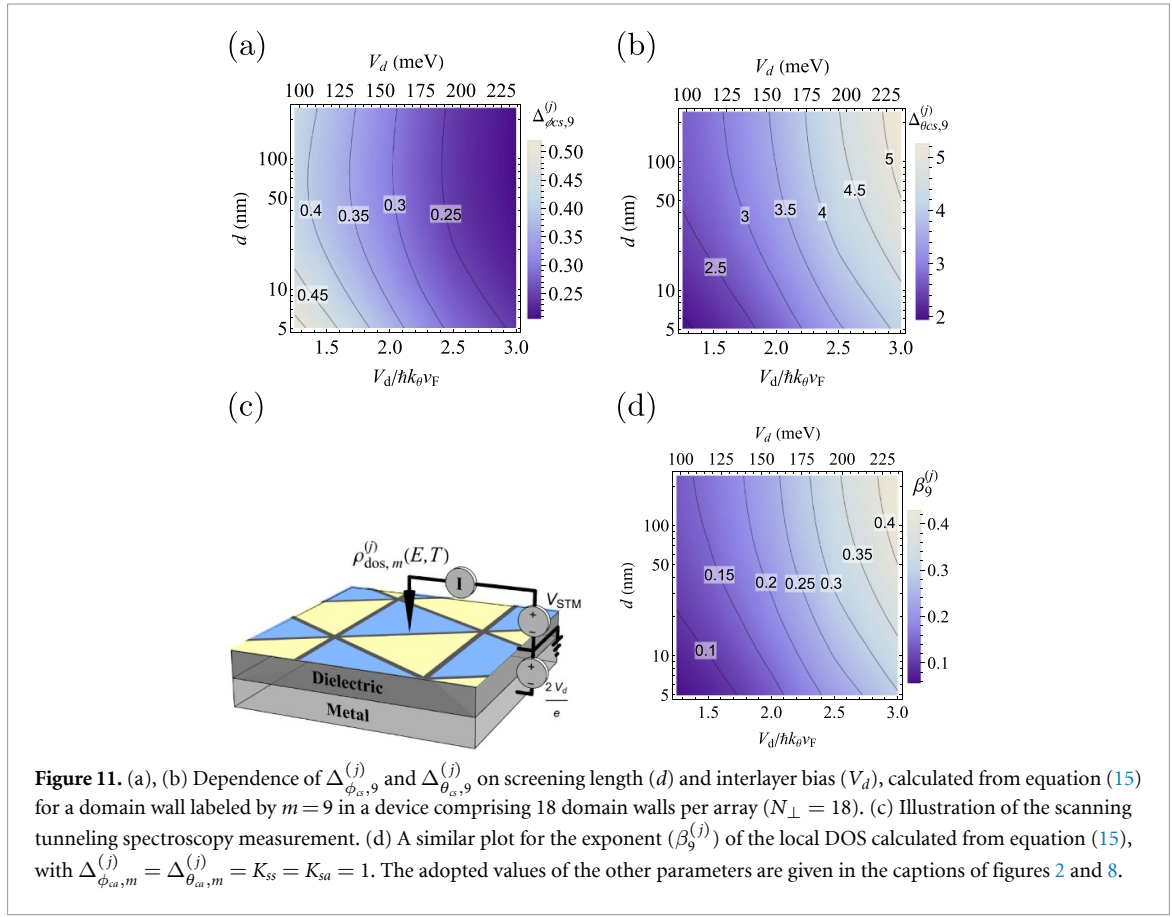

**Figure 11.** (a), (b) Dependence of $\Delta^{(j)}_{\phi_{cs},9}$ and $\Delta^{(j)}_{\theta_{cs},9}$ on screening length ($d$) and interlayer bias ($V_d$), calculated from equation (15) for a domain wall labeled by $m = 9$ in a device comprising 18 domain walls per array ($N_\perp = 18$). (c) Illustration of the scanning tunneling spectroscopy measurement. (d) A similar plot for the exponent ($\beta^{(j)}_9$) of the local DOS calculated from equation (15), with $\Delta^{(j)}_{\phi_{ca},m} = \Delta^{(j)}_{\theta_{ca},m} = K_{ss} = K_{sa} = 1$. The adopted values of the other parameters are given in the captions of figures 2 and 8.

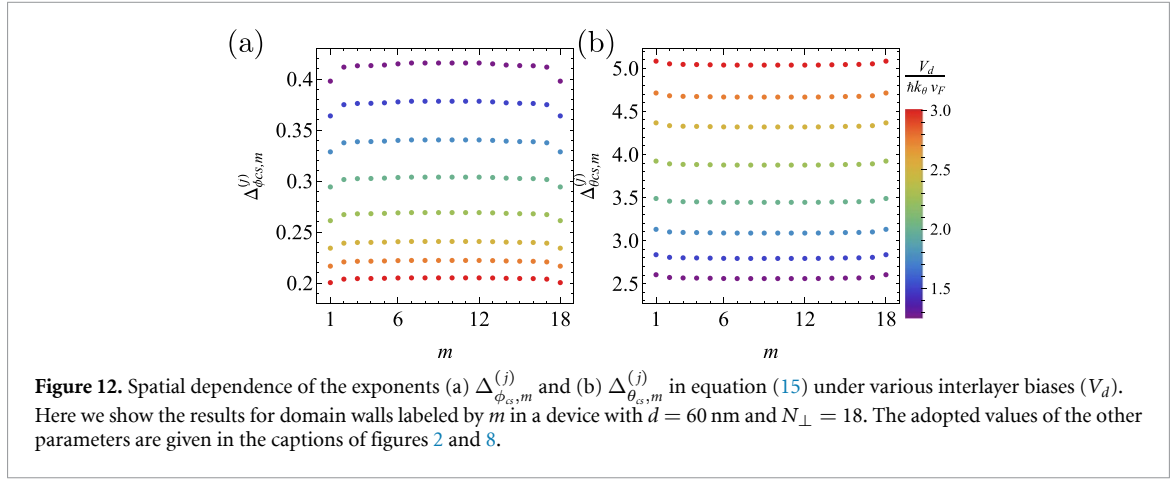

**Figure 12.** Spatial dependence of the exponents (a) $\Delta^{(j)}_{\phi_{cs},m}$ and (b) $\Delta^{(j)}_{\theta_{cs},m}$ in equation (15) under various interlayer biases ($V_d$). Here we show the results for domain walls labeled by $m$ in a device with $d = 60$ nm and $N_\perp = 18$. The adopted values of the other parameters are given in the captions of figures 2 and 8.

**5.2. Charge transport features at microscopic scales**
Here we discuss the charge transport features from three microscopic processes, which we illustrate in figure 10. To begin with, we consider an isolated impurity in the $m$th domain wall of the $j$th array, illustrated in figure 13(a). It can be modeled as a delta function $V_{\mathrm{imp}}(x) = v_b \delta(x)$, which induces the backscattering of the electrons within the domain wall,

$$
\begin{aligned}
H^{(j)}_{\mathrm{imp},m} &= \int dx\, V_{\mathrm{imp}}(x) \sum_{\delta\delta'\sigma} \left\{ \left[\psi^{(j)}_{R\delta\sigma,m}(x)\right]^\dagger \psi^{(j)}_{L\delta'\sigma,m}(x) \right.\\
&\quad\left. + \left[\psi^{(j)}_{L\delta'\sigma,m}(x)\right]^\dagger \psi^{(j)}_{R\delta\sigma,m}(x) \right\}, \\
&= \frac{2v_b}{\pi a}\sum_\delta \cos\left[\phi^j_{cs,m}(0) + \delta\phi^j_{ca,m}(0)\right] \\
&\quad\times \cos\left[\phi^j_{ss,m}(0) + \delta\phi^j_{sa,m}(0)\right] \\
&\quad + \frac{2v_b}{\pi a}\sum_\delta \cos\left[\phi^j_{cs,m}(0) - \delta\theta^j_{ca,m}(0)\right] \\
&\quad\times \cos\left[\phi^j_{ss,m}(0) - \delta\theta^j_{sa,m}(0)\right]. \quad (16)
\end{aligned}
$$

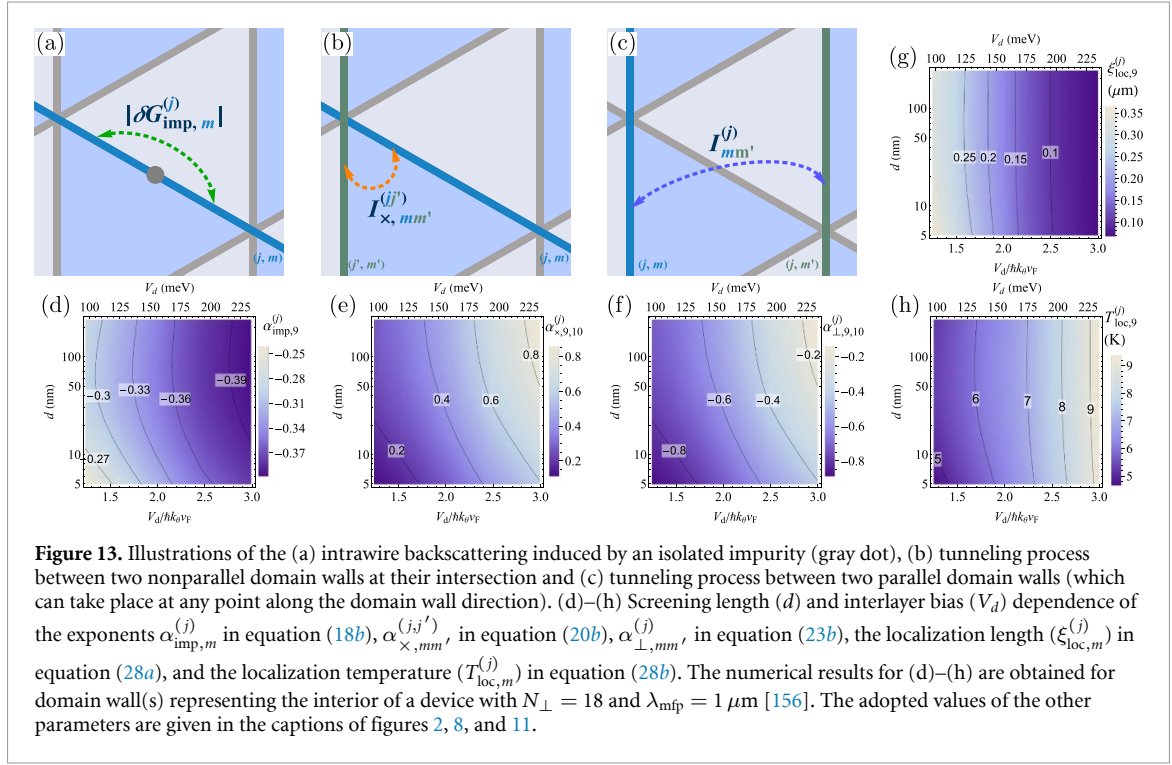

**Figure 13.** Illustrations of the (a) intrawire backscattering induced by an isolated impurity (gray dot), (b) tunneling process between two nonparallel domain walls at their intersection and (c) tunneling process between two parallel domain walls (which can take place at any point along the domain wall direction). (d)–(h) Screening length ($d$) and interlayer bias ($V_d$) dependence of the exponents $\alpha_{\mathrm{imp},m}^{(j)}$ in equation (18b), $\alpha_{\times,mm'}^{(j,j')}$, in equation (20b), $\alpha_{\perp,mm'}^{(j)}$ in equation (23b), the localization length ($\xi_{\mathrm{loc},m}^{(j)}$) in equation (28a), and the localization temperature ($T_{\mathrm{loc},m}^{(j)}$) in equation (28b). The numerical results for (d)–(h) are obtained for domain wall(s) representing the interior of a device with $N_\perp = 18$ and $\lambda_{\mathrm{mfp}} = 1\,\mu$m [156]. The adopted values of the other parameters are given in the captions of figures 2, 8, and 11.

In the above, we have intrabranch terms involving $\phi_{ca,m}^j$ field and interbranch terms involving $\theta_{ca,m}^j$ field. While their scaling behaviors are identical in our case with a noninteracting charge antisymmetric sector, in a more general situation the two terms might scale differently. We therefore focus on the scaling of the intrabranch terms, as these can be the most RG relevant term in either scenario. From equation (16), one can obtain the RG flow equation for the backscattering strength [4, 151, 152],

$$\frac{d\widetilde{v}_b}{dl} = \left[1 - \frac{1}{4}\left(\Delta_{\phi_{cs},m}^{(j)} + \Delta_{\phi_{ca},m}^{(j)} + K_{ss} + K_{sa}\right)\right]\widetilde{v}_b, \quad (17)$$

with $\widetilde{v}_b = v_b/\Delta_a$ and the dimensionless length scale $l$. It leads to a correction in the (differential) conductance,

$$|\delta G_{\mathrm{imp},m}^{(j)}| \propto \begin{cases} T^{\alpha_{\mathrm{imp},m}^{(j)}}, & \text{for } eV \ll k_B T, \\ V^{\alpha_{\mathrm{imp},m}^{(j)}}, & \text{for } eV \gg k_B T, \end{cases} \quad (18a)$$

$$\alpha_{\mathrm{imp},m}^{(j)} = \frac{\Delta_{\phi_{cs},m}^{(j)} + \Delta_{\phi_{ca},m}^{(j)} + K_{ss} + K_{sa}}{2} - 2. \quad (18b)$$

with an interaction-dependent exponent.

Next, assuming weak coupling between nonparallel domain walls, the intersections at the AA-stacking regions allow for tunneling process between electrons in two of the crossing domain walls, say, the $m$th domain wall of the array $j$ and the $m'$th domain wall of the array $j'$; see illustration in figure 13(b). To the leading order, the tunnel current between these domain walls can be obtained by generalizing the formula in one-dimensional systems [153, 154],

$$I_{\times,mm'}^{(j,j')} = \frac{e\left[t_{\times,mm'}^{(j,j')}\right]^2}{\hbar^2} \sum_\sigma \int_0^\infty dt \left\{ e^{-ieVt/\hbar}\left\langle \left[\psi_{\sigma,m}^{(j)}(t)\right]^\dagger \psi_{\sigma,m'}^{(j')}(t), \left[\psi_{\sigma,m'}^{(j')}(0)\right]^\dagger \psi_{\sigma,m}^{(j)}(0) \right\rangle \right.$$
$$\left. - e^{ieVt/\hbar}\left\langle \left[\psi_{\sigma,m'}^{(j')}(t)\right]^\dagger \psi_{\sigma,m}^{(j)}(t), \left[\psi_{\sigma,m}^{(j)}(0)\right]^\dagger \psi_{\sigma,m'}^{(j')}(0) \right\rangle \right\}, \quad (19)$$

with the field operators defined at the coordinates of the intersection. Here, the fields $\psi^{(j)}_{\sigma,m}$ and $\psi^{(j')}_{\sigma,m'}$ belong to the two domain walls involving the tunneling process, the tunnel amplitude $t^{(j,j')}_{\times,mm'}$, and the voltage difference $V$ between the two domain walls.

Generalizing the algebra in [155] for networks, we get the current–voltage curve for general $V$ and temperature $T$,

$$I^{(j,j')}_{\times,mm'} \propto T^{\alpha^{(j,j')}_{\times,mm'}+1} \sinh\left(\frac{eV}{2k_BT}\right) \times \left|\Gamma\left(1 + \frac{\alpha^{(j,j')}_{\times,mm'}}{2} + i\frac{eV}{2\pi k_BT}\right)\right|^2, \quad (20a)$$

$$\alpha^{(j,j')}_{\times,mm'} = \frac{1}{8}\left(\Delta^{(j)}_{\phi_{cs},m} + \Delta^{(j)}_{\theta_{cs},m} + \Delta^{(j)}_{\phi_{ca},m} + \Delta^{(j)}_{\theta_{ca},m} \right.$$
$$\left. + \Delta^{(j')}_{\phi_{cs},m'} + \Delta^{(j')}_{\theta_{cs},m'} + \Delta^{(j')}_{\phi_{ca},m'} + \Delta^{(j')}_{\theta_{ca},m'} \right.$$
$$\left. + 2K_{ss} + 2K_{sa} + \frac{2}{K_{ss}} + \frac{2}{K_{sa}}\right) - 2, \quad (20b)$$

which involves the local DOS of the two crossing domain walls. Alternatively to the current-bias curves, one can derive the RG flow equation for the tunnel amplitude,

$$\frac{d\tilde{t}^{(j,j')}_{\times,mm'}}{dl} = -\frac{\alpha^{(j,j')}_{\times,mm'}}{2}\tilde{t}^{(j,j')}_{\times,mm'}, \quad (21)$$

where $\tilde{t}^{(j,j')}_{\times,mm'} = t^{(j,j')}_{\times,mm'}/\Delta_a$ is the dimensionless coupling. From the RG flow equation, we obtain the differential tunneling conductance in the high-bias or high-temperature regime,

$$\frac{dI^{(j,j')}_{\times,mm'}}{dV} \propto \begin{cases} T^{\alpha^{(j,j')}_{\times,mm'}}, & \text{for } eV \ll k_BT, \\ V^{\alpha^{(j,j')}_{\times,mm'}}, & \text{for } eV \gg k_BT, \end{cases} \quad (22)$$

consistent with equation (20) in these limits.

Finally, we can also discuss the tunneling between two parallel wires within a given array $j$, as plotted in figure 13(c). In the high-$V$ or high-$T$ regime, we get the following power laws for the differential tunneling conductance,

$$\frac{dI^{(j)}_{\perp,mm'}}{dV} \propto \begin{cases} T^{\alpha^{(j)}_{\perp,mm'}}, & \text{for } eV \ll k_BT, \\ V^{\alpha^{(j)}_{\perp,mm'}}, & \text{for } eV \gg k_BT, \end{cases} \quad (23a)$$

$$\alpha^{(j)}_{\perp,mm'} = \frac{1}{8}\left(\Delta^{(j)}_{\phi_{cs},m} + \Delta^{(j)}_{\theta_{cs},m} + \Delta^{(j)}_{\phi_{ca},m} + \Delta^{(j)}_{\theta_{ca},m} \right.$$
$$\left. + \Delta^{(j)}_{\phi_{cs},m'} + \Delta^{(j)}_{\theta_{cs},m'} + \Delta^{(j)}_{\phi_{ca},m'} + \Delta^{(j)}_{\theta_{ca},m'} \right.$$
$$\left. + 2K_{ss} + 2K_{sa} + \frac{2}{K_{ss}} + \frac{2}{K_{sa}}\right) - 3, \quad (23b)$$

where we incorporate the tunneling taking place at any spatial points along the parallel domain walls [63, 157].

While a comprehensive analysis of the transport properties of the entire TBG device requires further exploration at the mesoscopic level, similar to previous studies in the single-particle picture [158–160], we can examine the exponents characterizing three distinct microscopic processes relevant to transport. To this end, we demonstrate how $\alpha^{(j)}_{\text{imp},m}$, $\alpha^{(j,j')}_{\times,mm'}$, and $\alpha^{(j)}_{\perp,mm'}$ vary with experimentally controllable parameters, as shown in figures 13(d)–(f), considering a noninteracting spin and charge antisymmetric sectors (that is, $\Delta^{(j)}_{\phi_{ca},m} = \Delta^{(j)}_{\theta_{ca},m} = K_{ss} = K_{sa} = 1$ for any $j$ and $m$). Once again, we observe that all the exponents are electrically tunable through the interlayer bias and screening length, with a more pronounced dependence on the former. This is consistent with the behaviors of $\Delta^{(j)}_{\phi_{cs},m}$ and $\Delta^{(j)}_{\theta_{cs},m}$ discussed in figure 11. Specifically, a larger interlayer bias leads to stronger interactions, resulting in a larger deviation of the exponents from their noninteracting values. Among the three exponents, the behavior of $\alpha^{(j)}_{\text{imp},m}$ is rather simple, as it exhibits a constant shift from $\Delta^{(j)}_{\phi_{cs},m}$ and thus follows an identical trend. Given that both exponents $\alpha^{(j,j')}_{\times,mm'}$ and $\alpha^{(j)}_{\perp,mm'}$ are linear functions of the sum $\Delta^{(j)}_{\phi_{cs},m} + \Delta^{(j)}_{\theta_{cs},m}$, they behave similarly. As previously mentioned, in an isotropic and (relatively) homogeneous network like the one studied here, both $\Delta^{(j)}_{\phi_{cs},m}$ and $\Delta^{(j)}_{\theta_{cs},m}$ show only a weak dependence on $m$.

As demonstrated in this work, electric transport at microscopic scales can be influenced by external parameters, which should consequently affect the transport properties of mesoscopic devices, in which the three explored microscopic processes take place. A detailed analysis of these mesoscopic networks should include the TLL-induced correlation effects discussed here, along with the magnitudes of the bare couplings, which depend on material or device specifics and are not included above. While such detailed analysis exceeds the scope of the present work, it is important to highlight that, as devices are scaled up, the localization effect induced by potential disorder may play a more pronounced role. This effect is particularly notable in one-dimensional interacting channels [4, 151, 152]. We will therefore explore this phenomenon in the following section.

### 5.3. Anderson localization of the network

Apart from an isolated impurity, one can also consider potential disorder in the $m$th domain wall of the $j$th array as a random potential $V_{\text{dis},m}^{(j)}$, and investigate the possibility of Anderson localization in the network. We further assume a random potential of the form,

$$\left\langle\!\!\left\langle V_{\text{dis},m}^{(j)}(x)\, V_{\text{dis},m'}^{(j')}(x') \right\rangle\!\!\right\rangle = D_{b,m}^{(j)} \delta_{jj'}\delta_{mm'}\delta(x-x'), \quad (24)$$

with the disorder average $\langle\!\langle \cdots \rangle\!\rangle$ and the disorder strength $D_{b,m}^{(j)} = \hbar^2 v_F^2/(2\pi\lambda_{\text{mfp}})$ related to the mean free path $\lambda_{\text{mfp}}$ of the sample. For estimation purposes, we employ the experimentally extracted mean free path from typical two-dimensional devices [156], since a corresponding estimate for the one-dimensional domain wall mode is not available. The actual disorder strength may thus be further decreased due to reduced intervalley scattering, similar to the helical channels in quantum spin Hall insulators [161–163]. As a side remark, while we examine a particular type of potential disorder locally coupled to the domain wall modes, other forms of disorder, such as local twist angle variation and distortion [114, 164, 165], might also influence the transport properties of the system.

The random potential considered here leads to elastic backscattering of electrons within the domain wall, which can be described as

$$H_{\text{dis},m}^{(j)} = \sum_{\delta\delta'\sigma}\int dx\, V_{\text{dis},m}^{(j)}(x)$$
$$\times \left[\psi_{R\delta\sigma,m}^{(j)}(x)\right]^\dagger \psi_{L\delta'\sigma,m}^{(j)}(x) + \text{H.c.}. \quad (25)$$

Upon bosonization and performing disorder average [4], it leads to a contribution to the action, as given in equation (B.3) in appendix B, which can be quantified with the effective backscattering strength,

$$\widetilde{D}_{b,m}^{(j)} = \frac{2D_{b,m}^{(j)} a}{\pi\hbar^2 v_{\text{dw}}^2}. \quad (26)$$

Similar to single wires [4, 151, 152], one can derive the following RG flow equation for the effective backscattering strength,

$$\frac{d\widetilde{D}_{b,m}^{(j)}}{dl} = \left[3 - \frac{1}{2}\left(\Delta_{\phi_{cs},m}^{(j)} + \Delta_{\phi_{ca},m}^{(j)} + K_{ss} + K_{sa}\right)\right]$$
$$\times \widetilde{D}_{b,m}^{(j)}. \quad (27)$$

The flow equation indicates the presence of a gapped phase when $\Delta_{\phi_{cs},m}^{(j)} + \Delta_{\phi_{ca},m}^{(j)} < 6 - K_{ss} - K_{sa} = 4$, indicating a possibility for the Anderson localization in the network for a system even with (weakly) attractive interactions.

Given that the renormalization of other parameters, such as the interaction strength and the velocity, contributes at higher orders [4], we neglect these sub-leading contributions. This allows us to compute the localization length and localization temperature directly from the RG flow equation in equation (27) and get

$$\xi_{\text{loc},m}^{(j)} = a\left(\frac{\pi\hbar^2 v_{\text{dw}}^2}{2D_{b,m}^{(j)} a}\right)^{1/\left[3-\left(\Delta_{\phi_{cs},m}^{(j)}+\Delta_{\phi_{ca},m}^{(j)}+K_{ss}+K_{sa}\right)/2\right]}, \quad (28a)$$

and

$$k_B T_{\text{loc},m}^{(j)} = \frac{\hbar v_{\text{dw}}}{\xi_{\text{loc},m}^{(j)}}, \quad (28b)$$

respectively.

In figures 13(g)–(h), we illustrate the influence of screening length and interlayer bias on the physical quantities. Remarkably, the results are a culmination of the analyses we have discussed so far. Specifically, they are influenced by multiple factors, including the velocity $v_{\text{dw}}$ of the domain wall modes presented in figure 4, as well as the interaction strength $U_{ee,n}$ summarized in figures 8 and 9. For the localization length, we observe a more pronounced dependence on the interlayer bias compared to the screening length. As the interlayer bias increases, it not only strengthens the interaction but also suppresses $v_{\text{dw}}$. These two effects collectively contribute to the localization length, as evidenced in equation (28a). The localization temperature, on the other hand, exhibits a quantitatively different behavior. Within the explored parameter range, we note only a mild variation in the localization temperature. This is attributed to a balanced competition between the effects of interaction strength and $v_{\text{dw}}$, as indicated by equation (28b), resulting in a region with weaker dependence on the external parameters.

Our findings indicate that, under typical conditions, the domain wall network is likely to transition into an Anderson insulator for domain walls longer than the order of 0.1 $\mu$m at temperatures of several kelvins or below. It is important to note that these results are contingent upon the actual disorder strength, which may vary from sample to sample. In our estimation, we have assumed a mean free path $\lambda_{\text{mfp}} = 1\,\mu$m to represent a sample with moderate mobility [156]. Considering cleaner samples or reduced intervalley backscattering strength, we anticipate a longer localization length and a correspondingly lower localization temperature for the domain wall network. Moreover, while the disorder strength influences these quantities, the RG relevance is entirely determined by equation (27) and thus the electron–electron interaction encoded by the parameters $\Delta_{\phi_{cs},m}^{(j)}$ and $\Delta_{\phi_{ca},m}^{(j)}$. Consequently, as one attempts to scale up the device, the system can ultimately reach the localization regime at sufficiently low temperatures, provided that the disorder strength remains finite.

## 6. Correlation functions in the domain wall network

As one of the key features of TLL systems, various correlation functions exhibit power-law decay [1–4]. These functions are defined by exponents highly dependent on the interaction strength and indicate the instability of the TLL towards various phases. In this section, we examine the instabilities of the correlated domain wall network towards the DW and SC phases. We analyze two scenarios, one without phonons in section 6.1 and the other with phonons in section 6.2. This investigation not only addresses the tendency to these instabilities but also illustrates the utility of the developed model in such analyses.

### 6.1. Correlation functions of purely electronic systems

As in previous works on TLL-like model [4], we start with DW correlation functions. To this end, we introduce the following two operators,

$$O_{c,m}^{(j)}(\boldsymbol{x}) = \sum_{\sigma\sigma'} \left[\psi_{\sigma,m}^{(j)}(\boldsymbol{x})\right]^\dagger \sigma_{\sigma\sigma'}^0 \psi_{\sigma',m}^{(j)}(\boldsymbol{x}), \quad (29a)$$

$$O_{s,m}^{\mu,(j)}(\boldsymbol{x}) = \sum_{\sigma\sigma'} \left[\psi_{\sigma,m}^{(j)}(\boldsymbol{x})\right]^\dagger \sigma_{\sigma\sigma'}^\mu \psi_{\sigma',m}^{(j)}(\boldsymbol{x}), \quad (29b)$$

for the charge and spin (number) density, respectively. In the above, we introduce the identity matrix $\sigma^0 = \mathbb{1}$, the Pauli matrix $\boldsymbol{\sigma}^\mu$ of the component $\mu \in \{x, y, z\}$, the coordinate $\boldsymbol{x} = (x, ut)$ with the velocity $u \approx v_{\rm dw}$.

The backscattering terms of the above can be defined as the CDW and SDW operators. Owing to the existence of two branches for each moving direction and spin, we obtain both intrabranch and interbranch terms, akin to those in carbon nanotubes [147, 148, 166]. We distinguish these two types of contributions and apply bosonization to the corresponding operators using equation (7). The explicit expressions of the operators are presented in equation (B.4) in appendix B. With these expressions, we evaluate the intrabranch and interbranch CDW and SDW correlation functions,

$$\left\langle \left[O_{\Omega\,\text{cdw},m}^{(j)}(\boldsymbol{x})\right]^\dagger O_{\Omega\,\text{cdw},m}^{(j)}(0) \right\rangle_{\rm ee}$$
$$= \sum_\delta \frac{e^{2ik_{F\delta,m}^{(j)}x}}{2(\pi a)^2} \left|\frac{a}{\boldsymbol{x}}\right|^{\zeta_{\Omega\,\text{cdw},m}^{(j)}}, \quad (30a)$$

$$\left\langle \left[O_{\hookleftarrow\,\text{cdw},m}^{(j)}(\boldsymbol{x})\right]^\dagger O_{\hookleftarrow\,\text{cdw},m}^{(j)}(0) \right\rangle_{\rm ee}$$
$$= \frac{e^{i\left(k_{F1,m}^{(j)}+k_{F2,m}^{(j)}\right)x}}{(\pi a)^2} \left|\frac{a}{\boldsymbol{x}}\right|^{\zeta_{\hookleftarrow\,\text{cdw},m}^{(j)}}, \quad (30b)$$

$$\left\langle \left[O_{\Omega\,\text{sdw},m}^{\mu,(j)}(\boldsymbol{x})\right]^\dagger O_{\Omega\,\text{sdw},m}^{\mu,(j)}(0) \right\rangle_{\rm ee}$$
$$= \sum_\delta \frac{e^{2ik_{F\delta,m}^{(j)}x}}{2(\pi a)^2} \left|\frac{a}{\boldsymbol{x}}\right|^{\zeta_{\Omega\,\text{sdw},m}^{\mu,(j)}}, \quad (30c)$$

$$\left\langle \left[O_{\hookleftarrow\,\text{sdw},m}^{\mu,(j)}(\boldsymbol{x})\right]^\dagger O_{\hookleftarrow\,\text{sdw},m}^{\mu,(j)}(0) \right\rangle_{\rm ee}$$
$$= \frac{e^{i\left(k_{F1,m}^{(j)}+k_{F2,m}^{(j)}\right)x}}{(\pi a)^2} \left|\frac{a}{\boldsymbol{x}}\right|^{\zeta_{\hookleftarrow\,\text{sdw},m}^{\mu,(j)}}, \quad (30d)$$

where the exponents corresponding to the CDW and $\mu$ component of the SDW are given in table 1, with the subscripts $\Omega$ and $\hookleftarrow$ denoting intrabranch and interbranch terms, respectively.

Similar to the DW correlation functions, we can also examine the pairing correlation functions. To this end, we introduce the following pairing operators for singlet superconductivity (SSC) and triplet superconductivity (TSC),

$$O_{\text{ssc},m}^{(j)}(\boldsymbol{x}) = \sum_{\sigma\sigma'}\sum_\delta \sigma \left[\psi_{R\delta\sigma,m}^{(j)}(\boldsymbol{x})\right]^\dagger \sigma_{\sigma\sigma'}^0 \left[\psi_{L\delta\overline{\sigma}',m}^{(j)}(\boldsymbol{x})\right]^\dagger, \quad (31a)$$

$$O_{\text{tsc},m}^{\mu,(j)}(\boldsymbol{x}) = \sum_{\sigma\sigma'}\sum_\delta \sigma \left[\psi_{R\delta\sigma,m}^{(j)}(\boldsymbol{x})\right]^\dagger \sigma_{\sigma\sigma'}^\mu \left[\psi_{L\delta\overline{\sigma}',m}^{(j)}(\boldsymbol{x})\right]^\dagger, \quad (31b)$$

which describe pairings with zero total momentum. Since interbranch pairing would involve Cooper pairs carrying nonzero momentum, we do not include it in the present analysis. With the explicit expressions given in equation (B.5) in appendix B, we obtain the pairing correlation functions,

$$\left\langle \left[O_{\text{ssc},m}^{(j)}(\boldsymbol{x})\right]^\dagger O_{\text{ssc},m}^{(j)}(0) \right\rangle_{\rm ee} = \frac{1}{(\pi a)^2}\left|\frac{a}{\boldsymbol{x}}\right|^{\zeta_{\text{ssc},m}^{(j)}}, \quad (32a)$$

$$\left\langle \left[O_{\text{tsc},m}^{\mu,(j)}(\boldsymbol{x})\right]^\dagger O_{\text{tsc},m}^{\mu,(j)}(0) \right\rangle_{\rm ee} = \frac{1}{(\pi a)^2}\left|\frac{a}{\boldsymbol{x}}\right|^{\zeta_{\text{tsc},m}^{\mu,(j)}}, \quad (32b)$$

where the exponents corresponding to SSC and the $\mu$ component of the TSC are given in table 1.

Similar to the single-wire case [4], each of these correlation functions is characterized by a power-law behavior. Owing to the correlations between the domain walls of the network, the quantities for a given domain wall are also influenced by others, thereby reflecting the correlated characteristics of the entire network. From the above correlation functions, we get the instability conditions listed in table 1.

To proceed, we determine the dominant instability tendencies toward various phases by identifying the slowest decaying correlation function(s) from equations (30) and (32) in the parameter regime where the corresponding condition(s) in table 1 are fulfilled. As evidenced in figures 11–12 and the fact that we have $\Delta^{(j)}_{\phi_{cs},m} < \Delta^{(j)}_{\theta_{cs},m}$ and $\Delta^{(j)}_{\phi_{cs},m} < 1$ for the parameter range under investigation, we expect the dominant instability of the correlated domain wall network is CDW and SDW along the domain walls, compared to SSC and TSC. Our model does not incorporate interactions or perturbations that break the spin rotational symmetry; hence, with $K_{ss} = K_{sa} = 1$, both the charge and spin sectors exhibit the same power-law behavior. We additionally examined the interwire correlation functions for both CDW and SDW (see equation (B.7)) and SC (see equation (B.9)). We found that these functions decay even more rapidly. Consequently, the explicit formulas are presented in appendix B, and we do not discuss them further in the main text.

**Table 1.** Various correlation types in the domain wall network, the exponents of the corresponding correlation functions, their explicit forms, and the instability conditions. The correlation types sharing identical exponents are grouped in the same row. The correlation functions are given in equations (30), (32), (35), and (38). The interaction-dependent parameters $\Delta^{(j)}_{\phi_{cP},m}$, $\Delta^{(j)}_{\theta_{cP},m}$, $\Gamma^{(j)}_{\phi_{cP},m}$, and $\Gamma^{(j)}_{\theta_{cP},m}$ are given in equations (15b), (15c), (36), and (39), respectively.

| Correlation type | Exponent | Explicit form | Instability condition |
|---|---|---|---|
| Purely electronic systems | | | |
| Intrabranch CDW and SDW$_z$ | $\zeta^{(j)}_{\circlearrowright\text{cdw},m} = \zeta^{z,(j)}_{\circlearrowright\text{sdw},m}$ | $(\Delta^{(j)}_{\phi_{cs},m} + \Delta^{(j)}_{\phi_{ca},m} + K_{ss} + K_{sa})/2$ | $\zeta^{(j)}_{\circlearrowright\text{cdw},m}, \zeta^{z,(j)}_{\circlearrowright\text{sdw},m} \leqslant 2$ |
| Intrabranch SDW$_x$ and SDW$_y$ | $\zeta^{x,(j)}_{\circlearrowright\text{sdw},m} = \zeta^{y,(j)}_{\circlearrowright\text{sdw},m}$ | $(\Delta^{(j)}_{\phi_{cs},m} + \Delta^{(j)}_{\phi_{ca},m} + 1/K_{ss} + 1/K_{sa})/2$ | $\zeta^{x,(j)}_{\circlearrowright\text{sdw},m}, \zeta^{y,(j)}_{\circlearrowright\text{sdw},m} \leqslant 2$ |
| Interbranch CDW and SDW$_z$ | $\zeta^{(j)}_{\hookrightarrow\text{cdw},m} = \zeta^{z,(j)}_{\hookrightarrow\text{sdw},m}$ | $(\Delta^{(j)}_{\phi_{cs},m} + \Delta^{(j)}_{\theta_{ca},m} + K_{ss} + 1/K_{sa})/2$ | $\zeta^{(j)}_{\hookrightarrow\text{cdw},m}, \zeta^{z}_{\hookrightarrow\text{sdw}} \leqslant 2$ |
| Interbranch SDW$_x$ and SDW$_y$ | $\zeta^{x,(j)}_{\hookrightarrow\text{sdw},m} = \zeta^{y,(j)}_{\hookrightarrow\text{sdw},m}$ | $(\Delta^{(j)}_{\phi_{cs},m} + \Delta^{(j)}_{\theta_{ca},m} + 1/K_{ss} + K_{sa})/2$ | $\zeta^{x,(j)}_{\hookrightarrow\text{sdw},m}, \zeta^{y,(j)}_{\hookrightarrow\text{sdw},m} \leqslant 2$ |
| Intrabranch SSC and TSC$_z$ | $\zeta^{(j)}_{\text{ssc},m} = \zeta^{z,(j)}_{\text{tsc},m}$ | $(\Delta^{(j)}_{\theta_{cs},m} + \Delta^{(j)}_{\theta_{ca},m} + K_{ss} + K_{sa})/2$ | $\zeta^{(j)}_{\text{ssc},m}, \zeta^{z,(j)}_{\text{tsc},m} \leqslant 2$ |
| Intrabranch TSC$_x$ and TSC$_y$ | $\zeta^{x,(j)}_{\text{tsc},m} = \zeta^{y,(j)}_{\text{tsc},m}$ | $(\Delta^{(j)}_{\theta_{cs},m} + \Delta^{(j)}_{\theta_{ca},m} + 1/K_{ss} + 1/K_{sa})/2$ | $\zeta^{x,(j)}_{\text{tsc},m}, \zeta^{y,(j)}_{\text{tsc},m} \leqslant 2$ |
| Electron-phonon-coupled systems | | | |
| Intrabranch CDW and SDW$_z$ | $\zeta^{(j)}_{\text{ph,cdw},m} = \zeta^{z,(j)}_{\text{ph,sdw},m}$ | $(\Gamma^{(j)}_{\phi_{cs},m} + \Delta^{(j)}_{\phi_{ca},m} + K_{ss} + K_{sa})/2$ | $\zeta^{(j)}_{\text{ph,cdw},m}, \zeta^{z,(j)}_{\text{ph,sdw},m} \leqslant 2$ |
| Intrabranch SDW$_x$ and SDW$_y$ | $\zeta^{x,(j)}_{\text{ph,sdw},m} = \zeta^{y,(j)}_{\text{ph,sdw},m}$ | $(\Gamma^{(j)}_{\phi_{cs},m} + \Delta^{(j)}_{\phi_{ca},m} + 1/K_{ss} + 1/K_{sa})/2$ | $\zeta^{x,(j)}_{\text{ph,sdw},m}, \zeta^{y,(j)}_{\text{ph,sdw},m} \leqslant 2$ |
| Intrabranch SSC and TSC$_z$ | $\zeta^{(j)}_{\text{ph,ssc},m} = \zeta^{z,(j)}_{\text{ph,tsc},m}$ | $(\Gamma^{(j)}_{\theta_{cs},m} + \Delta^{(j)}_{\theta_{ca},m} + K_{ss} + K_{sa})/2$ | $\zeta^{(j)}_{\text{ph,ssc},m}, \zeta^{z,(j)}_{\text{ph,tsc},m} \leqslant 2$ |
| Intrabranch TSC$_x$ and TSC$_y$ | $\zeta^{x,(j)}_{\text{ph,tsc},m} = \zeta^{y,(j)}_{\text{ph,tsc},m}$ | $(\Gamma^{(j)}_{\theta_{cs},m} + \Delta^{(j)}_{\theta_{ca},m} + 1/K_{ss} + 1/K_{sa})/2$ | $\zeta^{x,(j)}_{\text{ph,tsc},m}, \zeta^{y,(j)}_{\text{ph,tsc},m} \leqslant 2$ |

To conclude, the correlation functions can also be controlled through their dependence on the interaction-dependent exponents. While in the pure electronic system the instability towards DW phases always prevails that towards SC phases, below we demonstrate the appearance of the phonon-induced SC in the correlated domain wall network.

### 6.2. Correlation functions in the presence of phonons

In this section we consider the presence of phonons, which create a displacement field and thus coupled to the low-energy electronic modes (that is, the domain wall modes in our system). The Hamiltonian now contains three terms, $H_{ee} + H_{ph} + H_{ep}$, with the electron part in equation (9), the phonon part $H_{ph}$, and the coupling $H_{ep}$ between the two.

While the electron–phonon coupling in twisted structures is currently an important topic of research [77, 79, 167–172], a detailed microscopic model for this coupling in domain wall networks is not fully established. We therefore consider an effective model describing longitudinal acoustic phonons coupling to gapless modes that can move in one-dimensional channels [173, 174],

$$H_{\text{ph}} = \sum_{jm} \frac{1}{2\rho_a} \int dx \left[ \left(\Pi^j_{\text{ph},m}\right)^2 + \rho_a^2 c_{\text{ph}}^2 \left(\partial_x d^j_{\text{ph},m}\right)^2 \right], \quad (33)$$

with the effective mass density $\rho_a$ distributed along the domain wall, the phonon velocity $c_{\text{ph}}$, the conjugate field $\Pi^j_{\text{ph},m}$ of the displacement field $d^j_{\text{ph},m}$ generated by the phonons. The displacement field couples to the charge density in the domain walls, which in the bosonic language can be expressed as

$$H_{\text{ep}} = \sum_{jm} g_{\text{ep}} \int dx \left(\partial_x \phi^j_{cs,m}\right) \left(\partial_x d^j_{\text{ph},m}\right), \quad (34)$$

with the effective coupling $g_{\text{ep}}$ between the domain wall mode and the phonon. Here, since we assume the same coupling strengths for both branches of a given domain wall, the effective coupling only acts on the charge symmetric sector. In a more general case, however, phonons can also couple to the $\phi^j_{ca,m}$ field and thus affect the exponent in the charge antisymmetric sector.

In our analysis, we include the nonperturbative effects of electron–phonon coupling, thereby extending beyond the perturbative regime for $g_{\text{ep}}$, which we treat as a free parameter. It is essential to recognize that electron–phonon coupling may be significantly enhanced in the moiré structures, as highlighted by [79, 167, 170, 172]. Clearly, the incorporation of the phonon contributions leads to a breakdown in the duality between $\phi^j_{cs,m}$ and $\theta^j_{cs,m}$ typically existed in TLL [152]. This indicates a change in the properties of the domain wall modes, potentially leading to different electronic behavior of the network.

With the phonon-induced contribution, we derive the effective action in equations (C.1)–(C.3), as well as the correlators modified by phonons in equation (C.4). To investigate the correlation functions for various DW phases, we focus exclusively on intrabranch correlations here, motivated by the fact that interbranch correlations are, at best, as RG relevant as intrabranch correlations. With the details presented in appendix C, we compute the following correlation functions,

$$\left\langle \left[ O^{(j)}_{\Omega\,\mathrm{cdw},m}(\boldsymbol{x}) \right]^\dagger O^{(j)}_{\Omega\,\mathrm{cdw},m}(0) \right\rangle_{\mathrm{ee+ph}}$$
$$= \sum_\delta \frac{e^{2ik^{(j)}_{F\delta,m}x}}{2(\pi a)^2} \left| \frac{a}{\boldsymbol{x}} \right|^{\zeta^{(j)}_{\mathrm{ph,cdw},m}}, \quad (35a)$$

$$\left\langle \left[ O^{\mu,(j)}_{\Omega\,\mathrm{sdw},m}(\boldsymbol{x}) \right]^\dagger O^{\mu,(j)}_{\Omega\,\mathrm{sdw},m}(0) \right\rangle_{\mathrm{ee+ph}}$$
$$= \sum_\delta \frac{e^{2ik^{(j)}_{F\delta,m}x}}{2(\pi a)^2} \left| \frac{a}{\boldsymbol{x}} \right|^{\zeta^{\mu,(j)}_{\mathrm{ph,sdw},m}}, \quad (35b)$$

with $\langle\cdots\rangle_{\mathrm{ee+ph}}$ with respect to the action in equation (C.1). In the above, the exponents are given in table 1, with the phonon-modified parameters defined as

$$\Gamma^{(j)}_{\phi_{cs},m} = \sum_p \sqrt{\frac{\widetilde{U}^{(j)}_{\theta_{cs},p}}{\widetilde{U}^{(j)}_{\phi_{cs},p}}} \eta^j_{\phi,p} \left[ \left( \mathbf{M}^{(j)}_{\phi_{cs}} \right)_{p,m} \right]^2, \quad (36a)$$

$$\eta^j_{\phi,m} = u^j_m \left( \frac{\gamma^j_{\phi,+,m}}{u^j_{+,m}} + \frac{\gamma^j_{\phi,-,m}}{u^j_{-,m}} \right), \quad (36b)$$

$$\gamma^j_{\phi,\pm,m} = \pm \frac{\left(u^j_{\pm,m}\right)^2 - c^2_{\mathrm{ph}}}{\left(u^j_{+,m}\right)^2 - \left(u^j_{-,m}\right)^2}. \quad (36c)$$

Here, the phonon-induced modification stems from the factor $\eta^j_{\phi,m}$, which contains the following parameters obtained in the course of deriving the effective action in appendix C,

$$u^j_m = \frac{\sqrt{\widetilde{U}^{(j)}_{\phi_{cs},m} \widetilde{U}^{(j)}_{\theta_{cs},m}}}{\hbar}, \quad (37a)$$

$$v^j_{\mathrm{ep},m} = \left( \frac{g^2_{\mathrm{ep}}}{\hbar^2} \frac{\pi \widetilde{U}^{(j)}_{\theta_{cs},m}}{\rho_a} \right)^{1/4}, \quad (37b)$$

$$u^j_{\pm,m} = \left[ \frac{1}{2} \left( \left(u^j_m\right)^2 + c^2_{\mathrm{ph}} \right. \right.$$
$$\left. \left. \pm \sqrt{\left( \left(u^j_m\right)^2 - c^2_{\mathrm{ph}} \right)^2 + 4\left(v^j_{\mathrm{ep},m}\right)^4} \right) \right]^{1/2}. \quad (37c)$$

In the absence of electron–phonon coupling (that is, $v^j_{\mathrm{ep},m} \to 0$), the velocities $u^j_{\pm,m}$ correspond to the velocity $u^j_m$ of the charge symmetric mode in the $m$th domain wall of the $j$th array and that of the phonon. A nonzero electron–phonon coupling hybridizes the two, resulting in the renormalization of both velocities. This, combined with the characteristic features of TLL, can influence physical quantities via the exponents of various correlation functions discussed below. Similar to one-dimensional systems [173–177], in the limit of very strong electron–phonon coupling, there is a WB singularity where the velocity $u_{-,m}$ approaches zero, beyond which our model becomes unstable.

In the presence of phonons, we are particularly interested in the potential formation of SC phases within the domain wall network. This can be explored through the following SSC and TSC correlation functions,

$$\left\langle \left[ O^{(j)}_{\mathrm{ssc},m}(\boldsymbol{x}) \right]^\dagger O^{(j)}_{\mathrm{ssc},m}(0) \right\rangle_{\mathrm{ee+ph}} = \frac{1}{(\pi a)^2} \left| \frac{a}{\boldsymbol{x}} \right|^{\zeta^{(j)}_{\mathrm{ph,ssc},m}}, \quad (38a)$$

$$\left\langle \left[ O^{z,(j)}_{\mathrm{tsc},m}(\boldsymbol{x}) \right]^\dagger O^{z,(j)}_{\mathrm{tsc},m}(0) \right\rangle_{\mathrm{ee+ph}} = \frac{1}{(\pi a)^2} \left| \frac{a}{\boldsymbol{x}} \right|^{\zeta^{\mu,(j)}_{\mathrm{ph,tsc},m}}, \quad (38b)$$

with the exponents given in table 1 and the modified parameters,

$$\Gamma^{(j)}_{\theta_{cs},m} = \sum_p \sqrt{\frac{\widetilde{U}^{(j)}_{\phi_{cs},p}}{\widetilde{U}^{(j)}_{\theta_{cs},p}}} \eta^j_{\theta,p} \left[ \left( \mathbf{M}^{(j)}_{\theta_{cs}} \right)_{p,m} \right]^2 \quad (39a)$$

$$\eta^j_{\theta,m} = u^j_m \left( \frac{\gamma^j_{\theta,+,m}}{u^j_{+,m}} + \frac{\gamma^j_{\theta,-,m}}{u^j_{-,m}} \right), \quad (39b)$$

$$\gamma^j_{\theta,\pm,m} = \pm \frac{\left(u^j_{\pm,m}\right)^2 + \left(v^j_{\mathrm{ep},m}\right)^4 / \left(u^j_m\right)^2 - c^2_{\mathrm{ph}}}{\left(u^j_{+,m}\right)^2 - \left(u^j_{-,m}\right)^2}. \quad (39c)$$

The above results show that the pairing correlation in the network can be substantially influenced by the phonons. By comparing equations (36a) and (39a) with equations (15b) and (15c), it is evident that the phonon-induced effect is captured by the factor $\eta^j_{\phi,m}$ for the DW correlations and $\eta^j_{\theta,m}$ for the SC correlation. We therefore present their values in figure 14(a) and analyze how the effective electron–phonon coupling influences these factors, which are unity in the absence of phonons. A key observation is their opposite trend (that is, $\eta^j_{\phi,m} \geqslant 1$ and $\eta^j_{\theta,m} \leqslant 1$), implying that the phonons effectively induce attractive interaction between electrons in the domain wall network. The stronger the electron–phonon coupling is, the more pronounced the deviation of $\eta^j_{\phi,m}$ and $\eta^j_{\theta,m}$ is from unity. Beyond a critical value, a divergence occurs in the parameters, with $\eta^j_{\phi,m} \to \infty$ and $\eta^j_{\theta,m} \to 0$. Consequently, we identify the WB singularity at the coupling strength where the first of the modes reaches this divergence. Concerning the instability of the network, the $\eta$'s significantly influence the parameters $\Gamma^{(j)}_{\phi_{cs},m}$ and $\Gamma^{(j)}_{\theta_{cs},m}$ crucial in determining various correlation functions, as demonstrated in the inset. With a strong electron–phonon coupling, a lower $\eta^j_{\theta,m}$ value reduces the exponent in the pairing correlation function. Therefore, one can in general expect a slower decay of pairing correlations in the presence of phonons, indicating a stronger tendency towards SC.

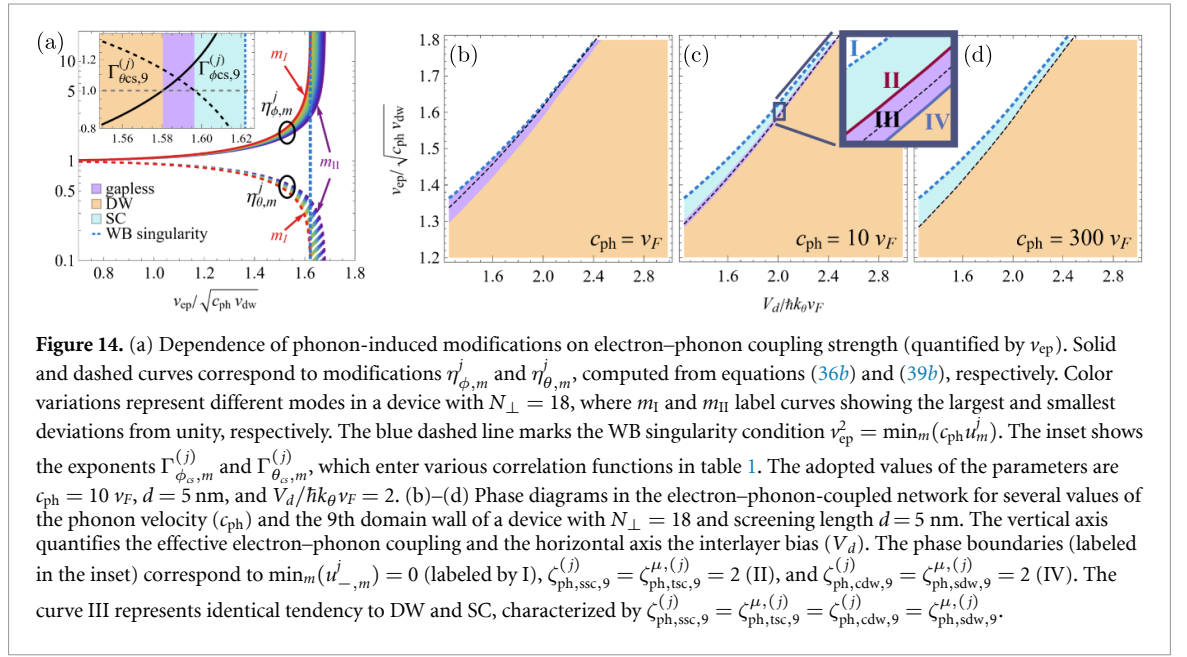

**Figure 14.** (a) Dependence of phonon-induced modifications on electron–phonon coupling strength (quantified by $v_{\text{ep}}$). Solid and dashed curves correspond to modifications $\eta^j_{\phi,m}$ and $\eta^j_{\theta,m}$, computed from equations (36b) and (39b), respectively. Color variations represent different modes in a device with $N_\perp = 18$, where $m_{\text{I}}$ and $m_{\text{II}}$ label curves showing the largest and smallest deviations from unity, respectively. The blue dashed line marks the WB singularity condition $v_{\text{ep}}^2 = \min_m(c_{\text{ph}} u^j_m)$. The inset shows the exponents $\Gamma^{(j)}_{\phi_{cs},m}$ and $\Gamma^{(j)}_{\theta_{cs},m}$, which enter various correlation functions in table 1. The adopted values of the parameters are $c_{\text{ph}} = 10\, v_F$, $d = 5$ nm, and $V_d/\hbar k_\theta v_F = 2$. (b)–(d) Phase diagrams in the electron–phonon-coupled network for several values of the phonon velocity ($c_{\text{ph}}$) and the 9th domain wall of a device with $N_\perp = 18$ and screening length $d = 5$ nm. The vertical axis quantifies the effective electron–phonon coupling and the horizontal axis the interlayer bias ($V_d$). The phase boundaries (labeled in the inset) correspond to $\min_m(u^j_{-,m}) = 0$ (labeled by I), $\zeta^{(j)}_{\text{ph,ssc},9} = \zeta^{\mu,(j)}_{\text{ph,tsc},9} = 2$ (II), and $\zeta^{(j)}_{\text{ph,cdw},9} = \zeta^{\mu,(j)}_{\text{ph,sdw},9} = 2$ (IV). The curve III represents identical tendency to DW and SC, characterized by $\zeta^{(j)}_{\text{ph,ssc},9} = \zeta^{\mu,(j)}_{\text{ph,tsc},9} = \zeta^{(j)}_{\text{ph,cdw},9} = \zeta^{\mu,(j)}_{\text{ph,sdw},9}$.

The correlation functions of the network obtained here share qualitative similarities with those in strictly one-dimensional channels; see, for instance, [4]. This similarity is evident in the power-law scaling, indicative of TLL characteristics, and in the general trend with the enhancement of DW and the suppression of SC due to repulsive electron–electron interactions. These findings align with the qualitative similarities between the one-dimensional systems and sTLL or csTLL [5–8]. Nonetheless, the inclusion of interwire correlations modifies the network instability on a quantitative level. For a quantitative analysis, we compute the exponents of the correlation functions and derive phase diagrams, the latter determined by the dominant instability exponents among those listed in table 1. It is important to note that the scope of analysis can be extended to include various phases induced by backscattering operators, such as the moiré correlated states discussed in [21]. Additionally, a unique aspect of the network here is the tunability of electron–electron interaction strength, adjustable through sample preparation and interlayer bias.

With the goal of providing a concrete estimation and inspired by the experimental setup described in [113], we consider a device with a short screening length, corresponding to a relatively strong screening effect. For our analysis below, the screened interaction strength can be subsequently modified by an interlayer bias. In addition, we examine three regimes characterized by the relative magnitude of phonon velocity to the Fermi velocity in a graphene monolayer. We use the latter quantity because the velocity of domain wall modes, being dependent on external parameters, does not serve well as an overall magnitude.

In figures 14(b)–(d), we present three types of phase diagrams, resulting from different values of the phonon velocity. As in figure 14(a), the WB singularity is indicated, which marks the strong electron–phonon coupling limit. Beyond this limit, the model in equation (9) becomes unstable. For velocities comparable to the graphene Fermi velocity, figure 14(b) shows a phase diagram featuring a DW phase and a correlated domain wall network phase. The latter phase corresponds to the parameter regime where none of the instability conditions listed in table 1 is fulfilled, and the network remains gapless, with its properties discussed in section 5. In [173, 174], the corresponding phase is identified as a metallic phase. It is also noteworthy that the phase diagram shown in figure 14(b) closely resembles the phase diagrams of systems that do not include phonons.

Interestingly, when the phonon velocity is sufficiently high, a new region emerges, as shown in figure 14(c). Specifically, with a sufficiently strong electron–phonon coupling strength, the dominant instability is SC at low $V_d$ values. At slightly higher $V_d$ values, a gapless network phase is observed, transitioning to a DW phase as electron–electron interactions increase at even higher $V_d$ values. Finally, an extremely high (and unrealistic) phonon velocity effectively suppresses the gapless network regime, as displayed in figure 14(d). Consequently, the region for SC becomes enlarged, as phonons mediate an instantaneous interaction between electrons.

In addition to the notable characteristics of electrically tunable network, which allows for *in-situ* driving of the system through phase transitions, we have several remarks. First, as has been identified in strictly one-dimensional interacting electronic systems [173–177], the WB singularity represents a phonon-driven instability in low-dimensional interacting electronic systems. Here, the location of this singularity in the parameter space, as defined in equation (37), is influenced by electron interaction and can thus be also electrically tuned, a feature absent in the previous literature [173–177]. Second, the phase diagrams are based on the exponent of a single domain wall in the interior of the device. As a result of the absence of translational invariance here, the exponents generally differ across the network, and the tendency towards SC or DW phases can vary. Consequently, the domain wall modes, including the two branches in each wall, progressively become unstable towards these phases in a nonuniform manner. We therefore expect a crossover, rather than a sharp transition in the network. Third, since the spin sectors are noninteracting ($K_{ss} = K_{sa} = 1$), the phase diagrams do not distinguish between spin-singlet and triplet phases. Introducing spin-dependent interactions, such as through proximity-induced spin–orbit coupling, could deviate from this specific case and further enrich these phase diagrams.

Before concluding this section, we note that the coupling strength between phonons and domain wall modes is a free parameter in our analysis, with its value not specified here. However, we observe that for a sufficiently strong coupling, the electronic properties can be adjusted to reach SC. As shown in figure 14(c), the range of SC broadens with decreasing electronic interaction strength. Additionally, as stated in section 3, our fitting procedure for the charge density is less accurate at very low $V_d$ values ($V_d \lesssim 1.2\hbar k_\theta v_F$) due to weaker confinement of domain wall modes in that regime. Nonetheless, based on the trends we have observed, it is reasonable to expect an extended SC phase at even lower $V_d$ values.

## 7. Discussion

In the present analysis, we show how the physical properties of the domain wall network in TBG can be tuned electrically. Motivated by the network observation in minimally TBG [38], in most parts of this work we consider a relatively small twist angle, specifically $\theta = 0.5°$, compared to the magic angle. This value corresponds to a larger effective hybridization parameter[3] $\alpha_{AB} \approx 1.4$. However, our findings are not limited to this particular value. As demonstrated in figures 4 and 6, we have explored a range of $\alpha_{AB} \in [1,2]$, corresponding to $0.35° \lesssim \theta \lesssim 0.7°$. Therefore, our analysis remains applicable across various parameter sets that result in the appearance of gapless modes within the domain walls.

We note that an even smaller twist can lead to a larger separation between the parallel domain walls, thereby reducing the interwire interaction strength. Conversely, a larger twist angle may bring the domain walls closer together, resulting in ill-defined one-dimensional channels. To ensure well-defined domain wall channels for the TLL description, the criterion $\frac{\rho_{2D}(r_{AB})}{\rho_{2D}(r_{dw})} \ll 1$ should be fulfilled. From the fitting function in equation (4), this criterion simplifies to $e^{-2\kappa_\perp} \ll 1$, a condition that is fulfilled across the entire range of investigation. To deduce the twist angle range satisfying the criterion, we examine a simplified expression from [30]. In our notation, it reads $e^{-w_{AB}\lambda_M/(2\pi\hbar v_F)} \ll 1$, indicating a twist angle on the order of $O(1°)$ or smaller. It is noteworthy to highlight that the analysis in [30] did not include relaxation effects. Given that stronger confinement of the domain wall modes is expected in the presence of relaxation, larger twist angles are possible for the feasible range of TLL description. Furthermore, while we focus on TBG in the present work, the procedure outlined above can be extended to other systems [61–64] in which similar one-dimensional channels have been identified.

From a broader perspective, we establish a mesoscopic network model from a single-particle model that describes interlayer hybridization at a more microscopic scale. This approach not only simplifies the subsequent analysis of correlated moiré systems with enlarged unit cells [31, 83], but also provides a broader implication of the network description in various settings. Namely, we present a systematic method for determining the interaction strength in networks formed by domain wall modes. Moreover, extending beyond the previous works of the moiré network [21, 67, 68, 72], here we explicitly incorporate the two branches for each moving direction and spin in a given domain wall. This consideration effectively establishes a mesoscopic network of coupled carbon nanotube TLLs.

Our approach further deviates from existing literature on sTLL or csTLL [5–8], which often rely on certain assumptions to build the bosonized model [21]. Specifically, these studies usually adopt a predefined form of the interaction. In contrast, we avoid presuming the forms of density–density and current–current interactions. Instead, we start directly with equation (9) and incorporate inputs from the outcomes derived from the single-particle continuum model in equation (1a). Additionally, we do not presume translational invariance perpendicular to the domain walls. Such invariance is not very realistic for a mesoscopic network with a limited number of parallel domain walls (fewer than 20 here), as opposed to the sTLL or csTLL description for CuO chains in typical cuprate samples.

---

[3] This choice is a balance for computational convenience, as an even larger $\alpha_{AB}$ would lead to steeper density profile, leading to a more time-consuming numerical procedure when calculating interaction strength in section 4.

In terms of observable features, our predictions include scaling behaviors in spectroscopic and transport measurements at microscopic scales. These behaviors are notably distinct from the topological edge modes of the quantum anomalous Hall states, where the scaling is determined by universal fractions [21]. In contrast to the topological nature of the edge modes, the scaling exponents here are directly influenced by the interaction strength, making them adjustable using the experimental parameters outlined above. This tunability consequently affects physical properties of mesoscopic devices, such as localization length and temperatures.

Furthermore, we examine the stability of this correlated domain wall network. We demonstrate that, while the screened Coulomb interaction in purely electronic systems favors the formation of CDW and SDW, the presence of longitudinal acoustic phonons can still induce pairing instability, driving the system towards SC instability and even the WB singularity. Our focus here has been on the properties of the network model as described by the *forward scattering* terms of the screened electron–electron interactions. Therefore, we have not covered those driven by backscattering operators, which can be analyzed through the (perturbative) RG procedure, as shown in [21, 67, 68, 72, 178]. Similarly, the $2k_F$ phonons [179–181] not considered here can also induce instability by coupling to the $2k_F$ component of the charge density. The analysis presented here can provide a more realistic view of the phase diagrams in those previous studies, where interaction strengths were introduced as free parameters.

Finally, our analysis indicates that twisted structures can manifest correlated phenomena by hosting spatially confined modes, without necessarily being tuned to magic angles. These modes are characterized by enhanced correlations and electrically tunable properties, forming the basis for an expanding theoretical literature employing bosonization [21, 66–68, 70–72]. In addition to TBG [30, 32, 35, 36, 38, 40, 45, 73], the explored one-dimensional modes also appear in a wider range of nanoscale systems, including twisted bilayer WTe$_2$ [63, 64, 74] and twisted trilayer graphene [61, 62]. Our work suggests that the exploration of domain wall network in twisted structures can open avenues for understanding and manipulating correlated phenomena in nanoscale systems.

## Data availability statement

All data that support the findings of this study are included within the article (and any supplementary files).

## Acknowledgment

We thank C-H Chang, P-Y Chang, C-K Chiu, Y-Z Chou, D Culcer, A Furusaki, T Hanaguri, Y-T Hsu, Y Kato, J Kishine, M Koshino, H Kusunose, T Otsuka, C K Shih, Y Yanase, B-J Yang, and N-C Yeh for stimulating discussions. We are grateful to N-C Yeh for pointing out the possibility of Anderson localization in the domain wall network. This work was supported financially by the National Science and Technology Council (NSTC), Taiwan through NSTC-112-2112-M-001-025-MY3. We are grateful to the long-term workshop YITP-T-23-01 held at Yukawa Institute for Theoretical Physics (YITP), Kyoto University, where a part of this work was carried out.

## Appendix A. Chemical potential dependence of the domain wall network properties

In this section, we examine the effect of the chemical potential on the density distribution of domain wall modes and, consequently, on the properties of the domain wall network in more details. Figure A1(a) presents the fitting parameters in equation (4) as a function of the chemical potential. Due to the particle-hole symmetry inherent in our single-particle Hamiltonian, the observed behavior exhibits symmetry for $\mu > 0$ and $\mu < 0$. We observe negligible changes for the parameters $\kappa_\perp$ and $c_{0\parallel}$, and a weak variation of approximately 10% for $\kappa_\parallel$ within the chemical potential range of the domain wall bands. Consequently, the spatial profile of the domain wall network is practically unchanged when the chemical potential stays within the energy bands near the charge neutrality point.

To further strengthen our statement, in figure A1(b), we present the computed values of the exponents $\Delta^{(j)}_{\phi_{cs},9}$ and $\Delta^{(j)}_{\theta_{cs},9}$. These parameters are crucial for determining the network's spectroscopic and transport properties, as well as its stability, discussed in the main text. Owing to the particle-hole symmetry, only the $\mu > 0$ side is shown. As expected, the negligible effect of chemical potential variations on the domain wall density leads to minor changes in these exponents, demonstrating the insensitivity of our results to the doping level.

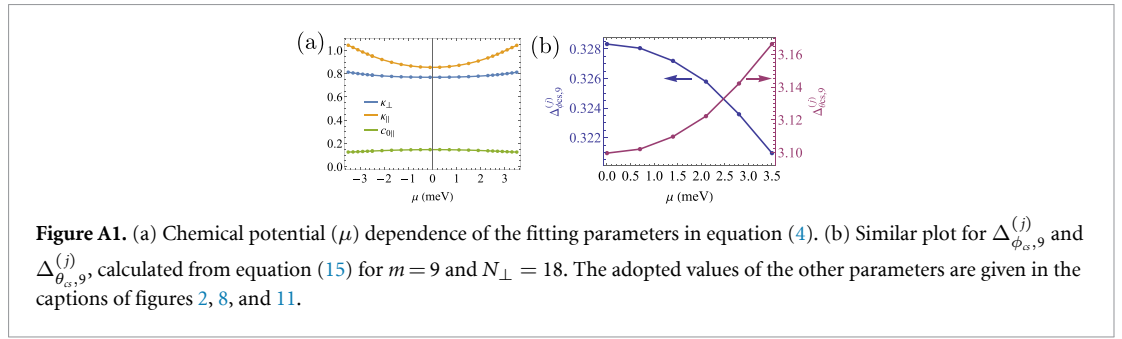

**Figure A1.** (a) Chemical potential ($\mu$) dependence of the fitting parameters in equation (4). (b) Similar plot for $\Delta^{(j)}_{\phi_{cs},9}$ and $\Delta^{(j)}_{\theta_{cs},9}$, calculated from equation (15) for $m = 9$ and $N_\perp = 18$. The adopted values of the other parameters are given in the captions of figures 2, 8, and 11.

## Appendix B. Details about the calculation in purely electronic systems

In this section, we present the details about our analysis when the phonons are absent. We begin by introducing the effective action and operators used to compute various correlation functions. In the diagonal basis, the electronic Hamiltonian (equation (9)) leads to the action,

$$\frac{S_{ee}}{\hbar} = \sum_{jm}\sum_{P} \int \frac{dx d\tau}{2\pi} \left[ -2i\left(\partial_x \widetilde{\theta}^j_{cP,m}\right)\left(\partial_\tau \widetilde{\phi}^j_{cP,m}\right) + \frac{\widetilde{U}^{(j)}_{\phi_{cP},m}}{\hbar}\left(\partial_x \widetilde{\phi}^j_{cP,m}\right)^2 + \frac{\widetilde{U}^{(j)}_{\theta_{cP},m}}{\hbar}\left(\partial_x \widetilde{\theta}^j_{cP,m}\right)^2 \right.$$
$$\left. -2i\left(\partial_x \theta^j_{sP,m}\right)\left(\partial_\tau \phi^j_{sP,m}\right) + \frac{u_{sP}}{K_{sP}}\left(\partial_x \phi^j_{sP,m}\right)^2 + u_{sP}K_{sP}\left(\partial_x \theta^j_{sP,m}\right)^2 \right], \quad \text{(B.1)}$$

with the imaginary time $\tau$. From this, we compute the following 'basic' correlation functions,

$$\left\langle \left[\widetilde{\phi}^j_{cs,m+n}(x,\tau) - \widetilde{\phi}^j_{cs,m}(0,0)\right]^2 \right\rangle_{ee} = \frac{1-\delta_{n0}}{2}\left(\sqrt{\frac{\widetilde{U}^{(j)}_{\theta_{cs},m+n}}{\widetilde{U}^{(j)}_{\phi_{cs},m+n}}} + \sqrt{\frac{\widetilde{U}^{(j)}_{\theta_{cs},m}}{\widetilde{U}^{(j)}_{\phi_{cs},m}}}\right)\ln\left|\frac{L}{a}\right| + \delta_{n0}\sqrt{\frac{\widetilde{U}^{(j)}_{\theta_{cs},m}}{\widetilde{U}^{(j)}_{\phi_{cs},m}}}\ln\left|\frac{\boldsymbol{x}}{a}\right|, \quad \text{(B.2a)}$$

$$\left\langle \left[\widetilde{\theta}^j_{cs,m+n}(x,\tau) - \widetilde{\theta}^j_{cs,m}(0,0)\right]^2 \right\rangle_{ee} = \frac{1-\delta_{n0}}{2}\left(\sqrt{\frac{\widetilde{U}^{(j)}_{\phi_{cs},m+n}}{\widetilde{U}^{(j)}_{\theta_{cs},m+n}}} + \sqrt{\frac{\widetilde{U}^{(j)}_{\phi_{cs},m}}{\widetilde{U}^{(j)}_{\theta_{cs},m}}}\right)\ln\left|\frac{L}{a}\right| + \delta_{n0}\sqrt{\frac{\widetilde{U}^{(j)}_{\phi_{cs},m}}{\widetilde{U}^{(j)}_{\theta_{cs},m}}}\ln\left|\frac{\boldsymbol{x}}{a}\right|, \quad \text{(B.2b)}$$

$$\left\langle \left[\widetilde{\phi}^j_{ca,m+n}(x,\tau) - \widetilde{\phi}^j_{ca,m}(0,0)\right]^2 \right\rangle_{ee} = \frac{1-\delta_{n0}}{2}\left(\sqrt{\frac{\widetilde{U}^{(j)}_{\theta_{ca},m+n}}{\widetilde{U}^{(j)}_{\phi_{ca},m+n}}} + \sqrt{\frac{\widetilde{U}^{(j)}_{\theta_{ca},m}}{\widetilde{U}^{(j)}_{\phi_{ca},m}}}\right)\ln\left|\frac{L}{a}\right| + \delta_{n0}\sqrt{\frac{\widetilde{U}^{(j)}_{\theta_{ca},m}}{\widetilde{U}^{(j)}_{\phi_{ca},m}}}\ln\left|\frac{\boldsymbol{x}}{a}\right|, \quad \text{(B.2c)}$$

$$\left\langle \left[\widetilde{\theta}^{j}_{ca,m+n}(x,\tau) - \widetilde{\theta}^{j}_{ca,m}(0,0)\right]^{2}\right\rangle_{ee} = \frac{1-\delta_{n0}}{2}\left(\sqrt{\frac{\widetilde{U}^{(j)}_{\phi_{ca},m+n}}{\widetilde{U}^{(j)}_{\theta_{ca},m+n}}} + \sqrt{\frac{\widetilde{U}^{(j)}_{\phi_{ca},m}}{\widetilde{U}^{(j)}_{\theta_{ca},m}}}\right)\ln\left|\frac{L}{a}\right| + \delta_{n0}\sqrt{\frac{\widetilde{U}^{(j)}_{\phi_{ca},m}}{\widetilde{U}^{(j)}_{\theta_{ca},m}}}\ln\left|\frac{x}{a}\right|, \tag{B.2d}$$

$$\left\langle \left[\phi^{j}_{ss,m+n}(x,\tau) - \phi^{j}_{ss,m}(0,0)\right]^{2}\right\rangle_{ee} = (1-\delta_{n0})K_{ss}\ln\left|\frac{L}{a}\right| + \delta_{n0}K_{ss}\ln\left|\frac{x}{a}\right|, \tag{B.2e}$$

$$\left\langle \left[\theta^{j}_{ss,m+n}(x,\tau) - \theta^{j}_{ss,m}(0,0)\right]^{2}\right\rangle_{ee} = (1-\delta_{n0})\frac{1}{K_{ss}}\ln\left|\frac{L}{a}\right| + \delta_{n0}\frac{1}{K_{ss}}\ln\left|\frac{x}{a}\right|, \tag{B.2f}$$

$$\left\langle \left[\phi^{j}_{sa,m+n}(x,\tau) - \phi^{j}_{sa,m}(0,0)\right]^{2}\right\rangle_{ee} = (1-\delta_{n0})K_{sa}\ln\left|\frac{L}{a}\right| + \delta_{n0}K_{sa}\ln\left|\frac{x}{a}\right|, \tag{B.2g}$$

$$\left\langle \left[\theta^{j}_{sa,m+n}(x,\tau) - \theta^{j}_{sa,m}(0,0)\right]^{2}\right\rangle_{ee} = (1-\delta_{n0})\frac{1}{K_{sa}}\ln\left|\frac{L}{a}\right| + \delta_{n0}\frac{1}{K_{sa}}\ln\left|\frac{x}{a}\right|, \tag{B.2h}$$

with the coordinate $\boldsymbol{x} = (x, u\tau)$ and $|\boldsymbol{x}| = \sqrt{x^{2}+(v_{\mathrm{dw}}|\tau|+a)^{2}}$. The formulas in equation (B.2) are used to compute various correlation functions in the purely electronic systems.

In the presence of potential disorder, we obtain a perturbation term in equation (25), which contains the intrabranch and interbranch contributions. In general, the two terms have different scalings. Since they contain mutually conjugate fields, they cannot be ordered simultaneously. We therefore focus on the intrabranch contribution, motivated by its RG relevance, in the main text. To proceed, we apply the replica method, perform the disorder average [4, 151, 152], and arrive at the following contribution to the action,

$$\frac{S^{(j)}_{\mathrm{dis},m}}{\hbar} = -\frac{\widetilde{D}^{(j)}_{b,m}v^{2}_{\mathrm{dw}}}{2\pi a^{3}}\sum_{rr'\delta}\int d\tau d\tau' dx \cos\left[\phi^{j;r}_{cs,m}(x,\tau) + \delta\phi^{j;r}_{ca,m}(x,\tau) - \phi^{j;r'}_{cs,m}(x,\tau') - \delta\phi^{j;r'}_{ca,m}(x,\tau')\right]$$
$$\times \cos\left[\phi^{j;r}_{ss,m}(x,\tau) + \delta\phi^{j;r}_{sa,m}(x,\tau)\right]\cos\left[\phi^{j;r'}_{ss,m}(x,\tau') + \delta\phi^{j;r'}_{sa,m}(x,\tau')\right], \tag{B.3}$$

where we have introduced a replica index $r, r'$ and the dimensionless coupling parameter $\widetilde{D}^{(j)}_{b,m}$ (see equation (26)). The above formula is used to derive the RG flow equation in equation (27) in the main text.

Here we give the expressions for the intrabranch ($\circlearrowright$) and interbranch ($\leftrightharpoons$) terms of the CDW and SDW operators,

$$O^{(j)}_{\circlearrowright\,\mathrm{cdw},m}(\boldsymbol{x}) = \frac{1}{\pi a}\sum_{\delta} e^{-2ik^{(j)}_{F\delta,m}x}e^{i\left(\phi^{j}_{cs,m}+\delta\phi^{j}_{ca,m}\right)}\cos\left(\phi^{j}_{ss,m}+\delta\phi^{j}_{sa,m}\right), \tag{B.4a}$$

$$O^{(j)}_{\leftrightharpoons\,\mathrm{cdw},m}(\boldsymbol{x}) = \frac{1}{\pi a}\sum_{\delta} e^{-i\left(k^{(j)}_{F1,m}+k^{(j)}_{F2,m}\right)x}e^{i\left(\phi^{j}_{cs,m}-\delta\theta^{j}_{ca,m}\right)}\cos\left(\phi^{j}_{ss,m}-\delta\theta^{j}_{sa,m}\right), \tag{B.4b}$$

$$O^{x,(j)}_{\circlearrowright\,\mathrm{sdw},m}(\boldsymbol{x}) = \frac{1}{\pi a}\sum_{\delta} e^{-2ik^{(j)}_{F\delta,m}x}e^{i\left(\phi^{j}_{cs,m}+\delta\phi^{j}_{ca,m}\right)}\cos\left(\theta^{j}_{ss,m}+\delta\theta^{j}_{sa,m}\right), \tag{B.4c}$$

$$O^{x,(j)}_{\leftrightharpoons\,\mathrm{sdw},m}(\boldsymbol{x}) = \frac{1}{\pi a}\sum_{\delta} e^{-i\left(k^{(j)}_{F1,m}+k^{(j)}_{F2,m}\right)x}e^{i\left(\phi^{j}_{cs,m}-\delta\theta^{j}_{ca,m}\right)}\cos\left(\theta^{j}_{ss,m}-\delta\phi^{j}_{sa,m}\right), \tag{B.4d}$$

$$O^{y,(j)}_{\circlearrowright\,\mathrm{sdw},m}(\boldsymbol{x}) = \frac{-1}{\pi a}\sum_{\delta} e^{-2ik^{(j)}_{F\delta,m}x}e^{i\left(\phi^{j}_{cs,m}+\delta\phi^{j}_{ca,m}\right)}\sin\left(\theta^{j}_{ss,m}+\delta\theta^{j}_{sa,m}\right), \tag{B.4e}$$

$$O^{y,(j)}_{\leftrightharpoons\,\mathrm{sdw},m}(\boldsymbol{x}) = \frac{1}{\pi a}\sum_{\delta} e^{-i\left(k^{(j)}_{F1,m}+k^{(j)}_{F2,m}\right)x}e^{i\left(\phi^{j}_{cs,m}-\delta\theta^{j}_{ca,m}\right)}\sin\left(\theta^{j}_{ss,m}-\delta\phi^{j}_{sa,m}\right), \tag{B.4f}$$

$$O^{z,(j)}_{\circlearrowright\,\mathrm{sdw},m}(\boldsymbol{x}) = \frac{i}{\pi a}\sum_{\delta} e^{-2ik^{(j)}_{F\delta,m}x}e^{i\left(\phi^{j}_{cs,m}+\delta\phi^{j}_{ca,m}\right)}\sin\left(\phi^{j}_{ss,m}+\delta\phi^{j}_{sa,m}\right), \tag{B.4g}$$

$$O^{z,(j)}_{\leftrightharpoons\,\mathrm{sdw},m}(\boldsymbol{x}) = \frac{i}{\pi a}\sum_{\delta} e^{-i\left(k^{(j)}_{F1,m}+k^{(j)}_{F2,m}\right)x}e^{i\left(\phi^{j}_{cs,m}-\delta\theta^{j}_{ca,m}\right)}\sin\left(\phi^{j}_{ss,m}-\delta\theta^{j}_{sa,m}\right), \tag{B.4h}$$

which, with equation (B.1), give the CDW and SDW correlation functions in equation (30) in the main text.

Similarly, working in the bosonic form, we have the following expressions for the pairing operators,

$$O_{\text{ssc},m}^{(j)}(x) = \frac{1}{\pi a} \sum_{\delta} e^{-i\left(\theta_{cs,m}^j + \delta\theta_{ca,m}^j\right)} \cos\left(\phi_{ss,m}^j + \delta\phi_{sa,m}^j\right), \tag{B.5a}$$

$$O_{\text{tsc},m}^{x,(j)}(x) = \frac{1}{\pi a} \sum_{\delta} e^{-i\left(\theta_{cs,m}^j + \delta\theta_{ca,m}^j\right)} \cos\left(\theta_{ss,m}^j + \delta\theta_{sa,m}^j\right), \tag{B.5b}$$

$$O_{\text{tsc},m}^{y,(j)}(x) = \frac{-1}{\pi a} \sum_{\delta} e^{-i\left(\theta_{cs,m}^j + \delta\theta_{ca,m}^j\right)} \sin\left(\theta_{ss,m}^j + \delta\theta_{sa,m}^j\right), \tag{B.5c}$$

$$O_{\text{tsc},m}^{z,(j)}(x) = \frac{i}{\pi a} \sum_{\delta} e^{-i\left(\theta_{cs,m}^j + \delta\theta_{ca,m}^j\right)} \sin\left(\phi_{ss,m}^j + \delta\phi_{sa,m}^j\right). \tag{B.5d}$$

which, with the action in equation (B.1), give the intrawire SSC and TSC correlation functions in equation (32) in the main text.

In addition to the intrawire correlations given above, we can generalize the DW operators to get their interwire counterparts in the network. Considering only the intrabranch contributions, we write down the operators,

$$O_{\text{cdw},\perp,n}^{(j)}(x) = \frac{1}{\pi a} \sum_{\delta} e^{-2i\left(k_{F\delta,m+n}^{(j)} - k_{F\delta,m}^{(j)}\right)x} e^{i\left(\phi_{cs,m+n}^j - \phi_{cs,m}^j + \delta\phi_{ca,m+n}^j - \delta\phi_{ca,m}^j\right)}$$
$$\times \cos\left[\left(\phi_{ss,m+n}^j - \phi_{ss,m}^j\right) + \delta\left(\phi_{sa,m+n}^j - \phi_{sa,m}^j\right)\right], \tag{B.6a}$$

$$O_{\text{sdw},\perp,n}^{x,(j)}(x) = \frac{1}{\pi a} \sum_{\delta} e^{-2i\left(k_{F\delta,m+n}^{(j)} - k_{F\delta,m}^{(j)}\right)x} e^{i\left(\phi_{cs,m+n}^j - \phi_{cs,m}^j + \delta\phi_{ca,m+n}^j - \delta\phi_{ca,m}^j\right)}$$
$$\times \cos\left[\left(\theta_{ss,m+n}^j - \theta_{ss,m}^j\right) + \delta\left(\theta_{sa,m+n}^j - \theta_{sa,m}^j\right)\right], \tag{B.6b}$$

$$O_{\text{sdw},\perp,n}^{y,(j)}(x) = \frac{-1}{\pi a} \sum_{\delta} e^{-2i\left(k_{F\delta,m+n}^{(j)} - k_{F\delta,m}^{(j)}\right)x} e^{i\left(\phi_{cs,m+n}^j - \phi_{cs,m}^j + \delta\phi_{ca,m+n}^j - \delta\phi_{ca,m}^j\right)}$$
$$\times \sin\left[\left(\theta_{ss,m+n}^j - \theta_{ss,m}^j\right) + \delta\left(\theta_{sa,m+n}^j - \theta_{sa,m}^j\right)\right], \tag{B.6c}$$

$$O_{\text{sdw},\perp,n}^{z,(j)}(x) = \frac{i}{\pi a} \sum_{\delta} e^{-2i\left(k_{F\delta,m+n}^{(j)} - k_{F\delta,m}^{(j)}\right)x} e^{i\left(\phi_{cs,m+n}^j - \phi_{cs,m}^j + \delta\phi_{ca,m+n}^j - \delta\phi_{ca,m}^j\right)}$$
$$\times \sin\left[\left(\phi_{ss,m+n}^j - \phi_{ss,m}^j\right) + \delta\left(\phi_{sa,m+n}^j - \phi_{sa,m}^j\right)\right], \tag{B.6d}$$

from which we compute the CDW and SDW correlations between the $n$th neighbor wires (for $n \neq 0$),

$$\left\langle \left[O_{\text{cdw},\perp,n}^{(j)}(x)\right]^\dagger O_{\text{cdw},\perp,n}^{(j)}(0) \right\rangle_{ee} \propto \frac{1}{(\pi a)^2} \sum_{\delta} e^{2i\left(k_{F\delta,m+n}^{(j)} - k_{F\delta,m}^{(j)}\right)x}$$
$$\times \left|\frac{a}{x}\right|^{\left(\Delta_{\phi_{cs},m}^{(j)} + \Delta_{\phi_{cs},m+n}^{(j)} + \Delta_{\phi_{ca},m}^{(j)} + \Delta_{\phi_{ca},m+n}^{(j)}\right)/2 - \overline{\Delta}_{\phi_{cs},m,n}^{(j)} - \overline{\Delta}_{\phi_{ca},m,n}^{(j)} + K_{ss} + K_{sa}}, \tag{B.7a}$$

$$\left\langle \left[O_{\text{sdw},\perp,n}^{x,(j)}(x)\right]^\dagger O_{\text{sdw},\perp,n}^{x,(j)}(0) \right\rangle_{ee} \propto \frac{1}{(\pi a)^2} \sum_{\delta} e^{2i\left(k_{F\delta,m+n}^{(j)} - k_{F\delta,m}^{(j)}\right)x}$$
$$\times \left|\frac{a}{x}\right|^{\left(\Delta_{\phi_{cs},m}^{(j)} + \Delta_{\phi_{cs},m+n}^{(j)} + \Delta_{\phi_{ca},m}^{(j)} + \Delta_{\phi_{ca},m+n}^{(j)}\right)/2 - \overline{\Delta}_{\phi_{cs},m,n}^{(j)} - \overline{\Delta}_{\phi_{ca},m,n}^{(j)} + 1/K_{ss} + 1/K_{sa}}, \tag{B.7b}$$

$$\left\langle \left[O_{\text{sdw},\perp,n}^{y,(j)}(x)\right]^\dagger O_{\text{sdw},\perp,n}^{y,(j)}(0) \right\rangle_{ee} \propto \frac{1}{(\pi a)^2} \sum_{\delta} e^{2i\left(k_{F\delta,m+n}^{(j)} - k_{F\delta,m}^{(j)}\right)x}$$
$$\times \left|\frac{a}{x}\right|^{\left(\Delta_{\phi_{cs},m}^{(j)} + \Delta_{\phi_{cs},m+n}^{(j)} + \Delta_{\phi_{ca},m}^{(j)} + \Delta_{\phi_{ca},m+n}^{(j)}\right)/2 - \overline{\Delta}_{\phi_{cs},m,n}^{(j)} - \overline{\Delta}_{\phi_{ca},m,n}^{(j)} + 1/K_{ss} + 1/K_{sa}}, \tag{B.7c}$$

$$\left\langle \left[O_{\text{sdw},\perp,n}^{z,(j)}(x)\right]^\dagger O_{\text{sdw},\perp,n}^{z,(j)}(0) \right\rangle_{ee} \propto \frac{1}{(\pi a)^2} \sum_{\delta} e^{2i\left(k_{F\delta,m+n}^{(j)} - k_{F\delta,m}^{(j)}\right)x}$$
$$\times \left|\frac{a}{x}\right|^{\left(\Delta_{\phi_{cs},m}^{(j)} + \Delta_{\phi_{cs},m+n}^{(j)} + \Delta_{\phi_{ca},m}^{(j)} + \Delta_{\phi_{ca},m+n}^{(j)}\right)/2 - \overline{\Delta}_{\phi_{cs},m,n}^{(j)} - \overline{\Delta}_{\phi_{ca},m,n}^{(j)} + K_{ss} + K_{sa}}, \tag{B.7d}$$

where we introduce another parameter,

$$\overline{\Delta}^{(j)}_{\phi_{cP},m,n} = \sum_p \sqrt{\frac{\widetilde{U}^{(j)}_{\theta_{cP},p}}{\widetilde{U}^{(j)}_{\phi_{cP},p}}} \left(\mathbf{M}^{(j)}_{\phi_{cP}}\right)_{p,m+n} \left(\mathbf{M}^{(j)}_{\phi_{cP}}\right)_{p,m}. \qquad (B.7e)$$

We note that $\overline{\Delta}^{(j)}_{\phi_{cP},m,n} \to \Delta^{(j)}_{\phi_{cP},m}$ when $n = 0$.

Similarly, we can generalize the pairing operators into their interwire counterparts,

$$O^{(j)}_{\text{ssc},\perp,n}(\boldsymbol{x}) = \frac{1}{\pi a} \sum_\delta e^{-i\left(\theta^j_{cs,m+n} - \theta^j_{cs,m} + \delta\theta^j_{ca,m+n} - \delta\theta^j_{ca,m}\right)} \cos\left[\left(\phi^j_{ss,m+n} - \phi^j_{ss,m}\right) + \delta\left(\phi^j_{sa,m+n} - \phi^j_{sa,m}\right)\right], \quad (B.8a)$$

$$O^{x,(j)}_{\text{tsc},\perp,n}(\boldsymbol{x}) = \frac{1}{\pi a} \sum_\delta e^{-i\left(\theta^j_{cs,m+n} - \theta^j_{cs,m} + \delta\theta^j_{ca,m+n} - \delta\theta^j_{ca,m}\right)} \cos\left[\left(\theta^j_{ss,m+n} - \theta^j_{ss,m}\right) + \delta\left(\theta^j_{sa,m+n} - \theta^j_{sa,m}\right)\right], \quad (B.8b)$$

$$O^{y,(j)}_{\text{tsc},\perp,n}(\boldsymbol{x}) = \frac{-1}{\pi a} \sum_\delta e^{-i\left(\theta^j_{cs,m+n} - \theta^j_{cs,m} + \delta\theta^j_{ca,m+n} - \delta\theta^j_{ca,m}\right)} \sin\left[\left(\theta^j_{ss,m+n} - \theta^j_{ss,m}\right) + \delta\left(\theta^j_{sa,m+n} - \theta^j_{sa,m}\right)\right], \quad (B.8c)$$

$$O^{z,(j)}_{\text{tsc},\perp,n}(\boldsymbol{x}) = \frac{i}{\pi a} \sum_\delta e^{-i\left(\theta^j_{cs,m+n} - \theta^j_{cs,m} + \delta\theta^j_{ca,m+n} - \delta\theta^j_{ca,m}\right)} \sin\left[\left(\phi^j_{ss,m+n} - \phi^j_{ss,m}\right) + \delta\left(\phi^j_{sa,m+n} - \phi^j_{sa,m}\right)\right], \quad (B.8d)$$

and compute the interwire SC correlations (for $n \neq 0$),

$$\left\langle \left[O^{(j)}_{\text{ssc},\perp,n}(\boldsymbol{x})\right]^\dagger O^{(j)}_{\text{ssc},\perp,n}(0)\right\rangle_{ee} \propto \frac{1}{(\pi a)^2} \left|\frac{a}{\boldsymbol{x}}\right|^{\left(\Delta^{(j)}_{\theta_{cs},m} + \Delta^{(j)}_{\theta_{cs},m+n} + \Delta^{(j)}_{\theta_{ca},m} + \Delta^{(j)}_{\theta_{ca},m+n}\right)/2 - \overline{\Delta}^{(j)}_{\theta_{cs},m,n} - \overline{\Delta}^{(j)}_{\theta_{ca},m,n} + K_{ss} + K_{sa}}, \qquad (B.9a)$$

$$\left\langle \left[O^{x,(j)}_{\text{tsc},\perp,n}(\boldsymbol{x})\right]^\dagger O^{x,(j)}_{\text{tsc},\perp,n}(0)\right\rangle_{ee} \propto \frac{1}{(\pi a)^2} \left|\frac{a}{\boldsymbol{x}}\right|^{\left(\Delta^{(j)}_{\theta_{cs},m} + \Delta^{(j)}_{\theta_{cs},m+n} + \Delta^{(j)}_{\theta_{ca},m} + \Delta^{(j)}_{\theta_{ca},m+n}\right)/2 - \overline{\Delta}^{(j)}_{\theta_{cs},m,n} - \overline{\Delta}^{(j)}_{\theta_{ca},m,n} + 1/K_{ss} + 1/K_{sa}}, \qquad (B.9b)$$

$$\left\langle \left[O^{y,(j)}_{\text{tsc},\perp,n}(\boldsymbol{x})\right]^\dagger O^{y,(j)}_{\text{tsc},\perp,n}(0)\right\rangle_{ee} \propto \frac{1}{(\pi a)^2} \left|\frac{a}{\boldsymbol{x}}\right|^{\left(\Delta^{(j)}_{\theta_{cs},m} + \Delta^{(j)}_{\theta_{cs},m+n} + \Delta^{(j)}_{\theta_{ca},m} + \Delta^{(j)}_{\theta_{ca},m+n}\right)/2 - \overline{\Delta}^{(j)}_{\theta_{cs},m,n} - \overline{\Delta}^{(j)}_{\theta_{ca},m,n} + 1/K_{ss} + 1/K_{sa}}, \qquad (B.9c)$$

$$\left\langle \left[O^{z,(j)}_{\text{tsc},\perp,n}(\boldsymbol{x})\right]^\dagger O^{z,(j)}_{\text{tsc},\perp,n}(0)\right\rangle_{ee} \propto \frac{1}{(\pi a)^2} \left|\frac{a}{\boldsymbol{x}}\right|^{\left(\Delta^{(j)}_{\theta_{cs},m} + \Delta^{(j)}_{\theta_{cs},m+n} + \Delta^{(j)}_{\theta_{ca},m} + \Delta^{(j)}_{\theta_{ca},m+n}\right)/2 - \overline{\Delta}^{(j)}_{\theta_{cs},m,n} - \overline{\Delta}^{(j)}_{\theta_{ca},m,n} + K_{ss} + K_{sa}}, \qquad (B.9d)$$

with another parameter,

$$\overline{\Delta}^{(j)}_{\theta_{cP},m,n} = \sum_p \sqrt{\frac{\widetilde{U}^{(j)}_{\phi_{cP},p}}{\widetilde{U}^{(j)}_{\theta_{cP},p}}} \left(\mathbf{M}^{(j)}_{\theta_{cP}}\right)_{p,m+n} \left(\mathbf{M}^{(j)}_{\theta_{cP}}\right)_{p,m}. \qquad (B.9e)$$

These interwire correlation functions decay faster than the intrawire correlations in equations (30) and (32); we therefore do not discuss them in the main text.

## Appendix C. Details about the calculation in the presence of phonons

In this section, we present the details about the calculation in the presence of phonons. Starting from equations (9), (33), and (34), we can perform straightforward algebra similar to [173, 174, 182] and derive the action,

$$\frac{S_{ee}}{\hbar} + \frac{S_{ph}}{\hbar} + \frac{S_{ep}}{\hbar} = \sum_{jm} \int \frac{dx d\tau}{2\pi} \left\{ \sum_P \left[ -2i\left(\partial_x \widetilde{\theta}^j_{cP,m}\right)\left(\partial_\tau \widetilde{\phi}^j_{cP,m}\right) + \frac{\widetilde{U}^{(j)}_{\phi_{cP},m}}{\hbar}\left(\partial_x \widetilde{\phi}^j_{cP,m}\right)^2 + \frac{\widetilde{U}^{(j)}_{\theta_{cP},m}}{\hbar}\left(\partial_x \widetilde{\theta}^j_{cP,m}\right)^2 \right. \right.$$

$$\left. - 2i\left(\partial_x \theta^j_{sP,m}\right)\left(\partial_\tau \phi^j_{sP,m}\right) + \frac{u_{sP}}{K_{sP}}\left(\partial_x \phi^j_{sP,m}\right)^2 + u_{sP} K_{sP}\left(\partial_x \theta^j_{sP,m}\right)^2 \right]$$

$$- \frac{2\pi i}{\hbar}\left(\Pi_{\text{ph},m}\right)\left(\partial_\tau d^j_{\text{ph},m}\right) + \frac{1}{2\rho_a \hbar}\left[\left(\Pi_{\text{ph},m}\right)^2 + \left(\partial_x d^j_{\text{ph},m}\right)^2\right]$$

$$\left. + \frac{g_{ep}}{\hbar}\sum_p \left(\mathbf{M}^{(j)}_{\phi_{cs}}\right)^{-1}_{m,p}\left(\partial_x \widetilde{\phi}^j_{cs,p}\right)\left(\partial_x d^j_{\text{ph},m}\right) \right\}, \qquad (C.1)$$

which includes electronic degrees of freedom, phonon fields, and their couplings. In the bosonic language, all these terms remain at most bilinear in the fields, still allowing for direct diagonalization. By integrating out the phonon fields, one can derive the effective action that incorporates phonon-mediated interaction in the following form,

$$\frac{S_{\text{ee+ph}}\left[\widetilde{\phi}^{j}_{cs,m}\right]}{\hbar} = \frac{1}{2\beta\hbar L}\sum_{jm}\sum_{\omega_n,q}\left[\frac{\omega_n^2 + \left(u_m^j\right)^2 q^2}{\pi \widetilde{U}^{(j)}_{\theta_{cs},m}/\hbar} - \frac{g_{\text{ep}}^2 q^4/(\rho_a\hbar)}{\omega_n^2 + c_{\text{ph}}^2 q^2}\right]\left|\widetilde{\phi}^{j}_{cs,m}(\omega_n,q)\right|^2, \quad \text{(C.2)}$$

and

$$\frac{S_{\text{ee+ph}}\left[\widetilde{\theta}^{j}_{cs,m}\right]}{\hbar} = \frac{1}{2\beta\hbar L}\sum_{jm}\sum_{\omega_n,q}\frac{\widetilde{U}^{(j)}_{\theta_{cs},m}}{\pi\hbar}\left[q^2 + \omega_n^2\frac{\omega_n^2 + c_{\text{ph}}^2 q^2}{\left(u_m^j\right)^2\left(\omega_n^2 + c_{\text{ph}}^2 q^2\right) - \left(v_{\text{ep},m}^j\right)^4 q^2}\right]\left|\widetilde{\theta}^{j}_{cs,m}(\omega_n,q)\right|^2, \quad \text{(C.3)}$$

with the parameter $u_m^j$ and $v_{\text{ep},m}^j$ given in equation (37). The latter depends on the domain wall index through $\widetilde{U}^{(j)}_{\theta_{cs},m}$. Since we have $\widetilde{U}^{(j)}_{\theta_{cP},m} = \hbar v_{\text{dw}}$ in our analysis, one may drop $j$ and $m$ in $v_{\text{ep},m}^j$ for simplicity.

With the effective action, we compute the following correlation functions under the influence of the phonons,

$$\left\langle\left[\widetilde{\phi}^{j}_{cs,m}(x,\tau) - \widetilde{\phi}^{j}_{cs,m}(0,0)\right]^2\right\rangle_{ee+ph} = \sqrt{\frac{\widetilde{U}^{(j)}_{\theta_{cs},m}}{\widetilde{U}^{(j)}_{\phi_{cs},m}}}\eta^{j}_{\phi,m}\ln\left|\frac{\boldsymbol{x}}{a}\right|, \quad \text{(C.4a)}$$

$$\left\langle\left[\widetilde{\theta}^{j}_{cs,m}(x,\tau) - \widetilde{\theta}^{j}_{cs,m}(0,0)\right]^2\right\rangle_{ee+ph} = \sqrt{\frac{\widetilde{U}^{(j)}_{\phi_{cs},m}}{\widetilde{U}^{(j)}_{\theta_{cs},m}}}\eta^{j}_{\theta,m}\ln\left|\frac{\boldsymbol{x}}{a}\right|, \quad \text{(C.4b)}$$

with the dimensionless parameters $\eta^{j}_{\phi,m}$ and $\eta^{j}_{\theta,m}$ introduced in equations (36) and (39), respectively.

In addition, the interwire CDW and SDW correlations in the presence of phonons take the form of equation (B.7) upon the replacement of $\Delta^{(j)}_{\phi_{cs},m} \to \Gamma^{(j)}_{\phi_{cs},m}$ with equation (36) and $\overline{\Delta}^{(j)}_{\phi_{cs},m,n} \to \overline{\Gamma}^{(j)}_{\phi_{cs},m,n}$ with

$$\overline{\Gamma}^{(j)}_{\phi_{cs},m,n} = \sum_p \sqrt{\frac{\widetilde{U}^{(j)}_{\theta_{cs},p}}{\widetilde{U}^{(j)}_{\phi_{cs},p}}}\eta^{j}_{\phi,p}\left(\mathbf{M}^{(j)}_{\phi_{cs}}\right)_{p,m+n}\left(\mathbf{M}^{(j)}_{\phi_{cs}}\right)_{p,m}. \quad \text{(C.5)}$$

Finally, the interwire pairing correlations in the presence of phonons take the form of equation (B.9) upon the replacement of $\Delta^{(j)}_{\theta_{cs},m} \to \Gamma^{(j)}_{\theta_{cs},m}$ with equation (39) and $\overline{\Delta}^{(j)}_{\theta_{cs},m,n} \to \overline{\Gamma}^{(j)}_{\theta_{cs},m,n}$ with

$$\overline{\Gamma}^{(j)}_{\theta_{cs},m,n} = \sum_p \sqrt{\frac{\widetilde{U}^{(j)}_{\phi_{cs},p}}{\widetilde{U}^{(j)}_{\theta_{cs},p}}}\eta^{j}_{\theta,p}\left(\mathbf{M}^{(j)}_{\theta_{cs}}\right)_{p,m+n}\left(\mathbf{M}^{(j)}_{\theta_{cs}}\right)_{p,m}. \quad \text{(C.6)}$$

## ORCID iDs

Hao-Chien Wang  https://orcid.org/0009-0008-9162-7584
Chen-Hsuan Hsu  https://orcid.org/0000-0003-0963-538X